\documentclass[journal,twoside]{IEEEtran}

\makeatletter
\def\ps@headings{%
\def\@oddhead{\mbox{}\scriptsize\rightmark \hfil \thepage}%
\def\@evenhead{\scriptsize\thepage \hfil \leftmark\mbox{}}%
\def\@oddfoot{}%
\def\@evenfoot{}}
\makeatother
\usepackage{graphicx}
\usepackage{subcaption}
\usepackage{amssymb,amsmath}
\usepackage [noadjust]{cite}
\usepackage{caption}
\usepackage{booktabs}
\usepackage{multirow}
\usepackage{arydshln}
\usepackage{cases}
\usepackage{mathabx}
\usepackage[linesnumbered,ruled,vlined]{algorithm2e}

\usepackage[rawfloats=true]{floatrow} 
\restylefloat{figure}
\usepackage{lscape}

\usepackage{amssymb}
\usepackage{pifont}
\usepackage{color, colortbl}

\newcommand{\M}{\mathcal{M}}

\newcommand{\A}{\mathcal{A}}
\newcommand{\B}{\mathcal{B}}
\newcommand{\s}{\mathcal{S}}

\newcommand{\X}{\mathcal{X}}

\newcommand{\tPrm}{t^{\prime}}

\newcommand{\Boldmu}{\mbox{\boldmath{$\mu$}}}

\newcommand{\Boldtheta}{\mbox{\boldmath{$\theta$}}}
\newcommand{\Boldeps}{\mbox{\boldmath{$\epsilon$}}}

\newtheorem {theorem*}{Theorem (Brouwer's Fixed Point Theorem)}

\begin{document}

%
\title{Generative AI for the Optimization of Next-Generation Wireless Networks: Basics, State-of-the-Art, and Open Challenges}

\author{Fahime Khoramnejad and Ekram Hossain,~\IEEEmembership{Fellow,~IEEE}

\thanks{Fahime~Khoramnejad and  Ekram Hossain are with the Department of Electrical and Computer Engineering, University of Manitoba, Winnipeg, Manitoba, Canada (e-mail: fahimeh.khoramnejad@umanitoba.ca and ekram.hossain@umanitoba.ca).} }

	{}

\maketitle
%
\begin{abstract}
Next-generation (xG) wireless networks, with their complex and dynamic nature, present significant challenges to using traditional optimization techniques. Generative AI (GAI) emerges as a powerful tool due to its unique strengths. Unlike traditional optimization techniques and other machine learning methods, GAI excels at learning from real-world network data, capturing its intricacies. This enables safe, offline exploration of various configurations and generation of diverse, unseen scenarios, empowering proactive, data-driven exploration and optimization for xG networks. Additionally, GAI's scalability makes it ideal for large-scale xG networks. This paper surveys how GAI-based models unlock optimization opportunities in xG wireless networks. We begin by providing a review of GAI models and some of the major communication paradigms of xG (e.g., 6G)  wireless networks. We then delve into  exploring how GAI can be used to improve resource allocation and enhance overall network performance. Additionally, we briefly review the networking requirements for supporting GAI applications in xG wireless networks. The paper further discusses the key challenges and future research directions in leveraging GAI for network optimization. Finally, a case study demonstrates the application of a diffusion-based GAI model for load balancing, carrier aggregation, and backhauling optimization in non-terrestrial networks, a core technology of xG networks. This case study serves as a practical example of how the combination of reinforcement learning and GAI can be implemented to address real-world network optimization problems.
\end{abstract}

    

\section{Introduction}
Network optimization entails enhancing the performance and efficiency of communication networks through various strategies and technologies. These methods aim to improve data transfer speeds, reduce latency, increase network capacity, and ensure reliable and secure connectivity. In the realm of next-generation (xG) wireless communications, such as 5G and 6G, significant challenges include managing increased network complexity and scale, dynamically adapting to service demands, achieving energy and resource sustainability, maintaining quality of service and user experience across diverse applications, and enhancing security to counter evolving threats. The integration of heterogeneous networks also presents a considerable challenge in maintaining seamless connectivity across various network types. To address these challenges and ensure seamless user experiences in the future, xG wireless networks require innovative solutions. Artificial intelligence (AI) offers significant potential in this area. 

\begin{figure}[t!]
  	\centering
  	\includegraphics[width=150pt]{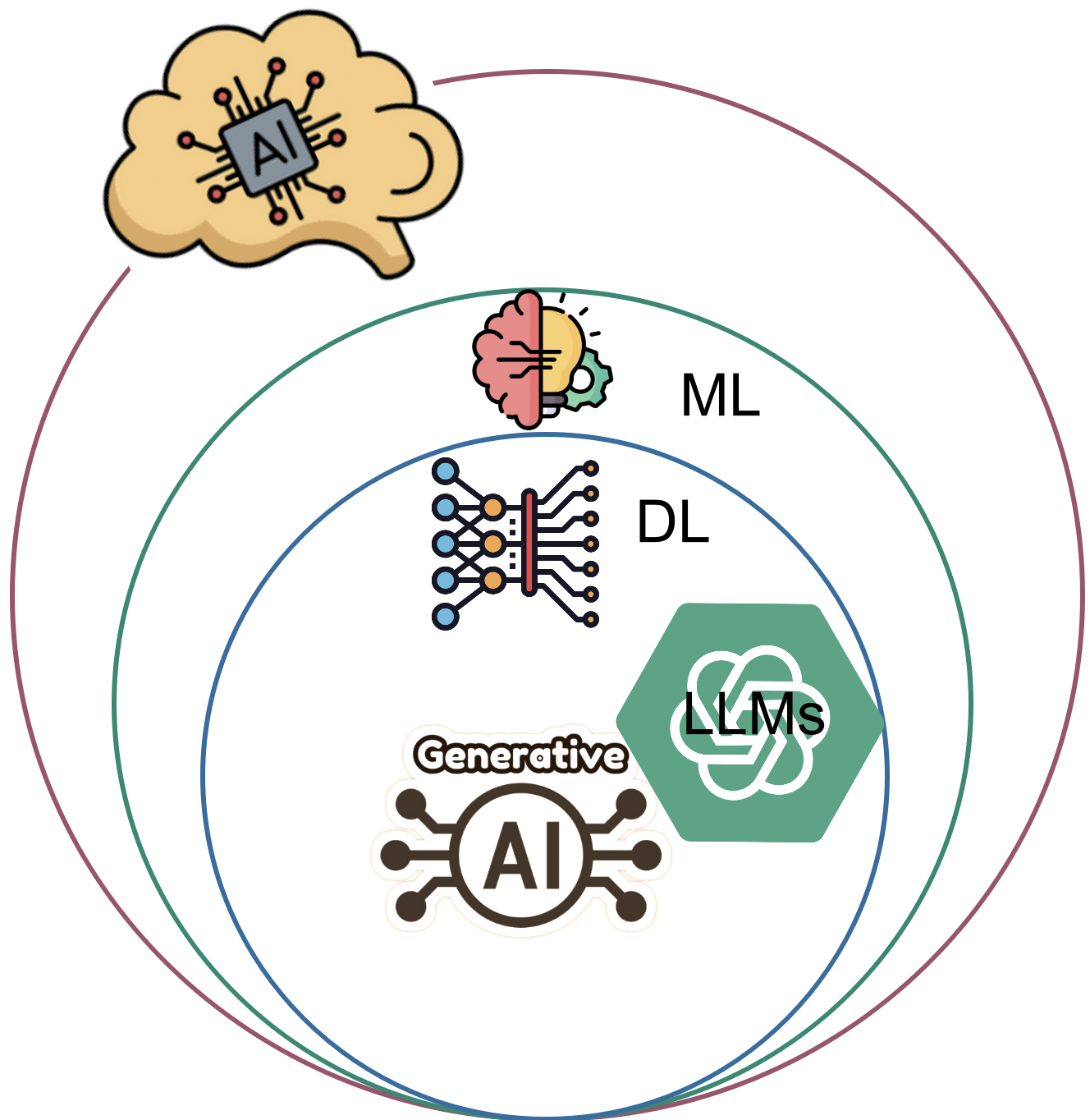}
  	\caption{AI, machine learning, deep learning, and GAI: A hierarchical overview.}
  	\label{fig_AI}
  \end{figure}

AI refers to the capability of a machine to imitate intelligent human behavior. AI systems aim to perform tasks that typically require human intelligence, such as understanding natural language, recognizing patterns, solving problems, and making decisions by combining computer science, data analysis,  and more. This enables the machine to perform tasks intelligently. 
{As illustrated in Fig.~\ref{fig_AI}, AI encompasses machine learning (ML), deep learning (DL), generative AI (GAI), and large language models (LLMs). ML allows machines to learn from data, while DL uses neural networks with multiple layers to learn from vast datasets. GAI is a specialized subset of AI that excels in creating new content from learned data patterns, such as text, images, videos, or code. This capability enables GAI systems to generate data instances that resemble, but are not exact copies of, their training examples. LLMs are a further specialization within GAI and DL, designed for natural language processing tasks. LLMs leverage the principles of DL and the generative capabilities of AI to understand and generate human-like text and perform a variety of language-based tasks such as translation and summarization.} 

The integration of AI  such as DL and reinforcement learning (RL) into LTE and 5G networks substantially advances their management and optimization. As the current generation of wireless networks evolves into xG networks (e.g., 6G networks), GAI would play a pivotal role in optimizing these advanced systems, ensuring they meet the increasing demands for speed, connectivity, and reliability in an ever-connected world. Different GAI models include generative adversarial networks (GANs), generative flow networks (GFlowNets), generative diffusion models (GDMs), variational autoencoders (VAEs), and autoencoders. In GANs, two neural networks, a generator and a discriminator, compete against each other to produce realistic data over time. GFlowNets use invertible mappings between data distributions and latent spaces to generate new data samples, learning to model the probability distribution of the training data directly. By iteratively refining random noise through a learned diffusion process, GDMs progressively transform simple noise into complex data that resembles the training set. Autoencoders, which are used for data generation and dimensionality reduction, consist of an encoder that maps input data to a latent representation and a decoder that reconstructs the data from this representation.

GAI's ability to rapidly produce unique and customizable content significantly boosts productivity across various industries by automating creative processes and supporting critical decision-making in fields like design and scientific research, including drug discovery. In wireless communications, GAI improves network design and optimization by simulating and testing complex network scenarios, which enhances signal coverage, facilitates capacity planning, and supports the deployment of advanced technologies. Consequently, GAI can promote more efficient and reliable communication services. The following subsections will discuss the basics of the GAI models and explore how GAI models further contribute to advancing network capabilities.

\subsection{Key Technologies Shaping xG Wireless Networks}
The landscape of wireless communication is enriched by the emergence of novel technologies. Some of these key technologies will be explained in detail below.

\vspace{0.2cm}
\noindent
\textbf{Artificial intelligence-generated content (AIGC):} It refers to any type of content created by AI models, including text, images, audio, and video. At the core of AIGC lies machine learning (ML). AI models are trained on massive datasets relevant to the content they generate. During training, the model identifies patterns and relationships within the data, learning the underlying structure and style. This knowledge allows the model to generate entirely new content that resembles the training data, but with an original twist. AIGC has a wide range of applications, including content creation, product design, art and music, and education. It can automate content creation, saving time and resources, and explore a wider range of creative possibilities than humans alone.

One exciting application of AIGC is its use on mobile devices, known as mobile AIGC. Mobile AIGC leverages advancements in mobile hardware and software. Powerful processors, efficient memory utilization, and on-device machine learning capabilities allow for running AIGC models directly on smartphones and other mobile devices. This eliminates the need for constant internet connectivity and cloud processing, making mobile AIGC more accessible and user-friendly. Unlike traditional AIGC focused on large-scale content creation, mobile AIGC emphasizes user-centric applications that can be readily used on the go. This could include real-time content creation, image and video editing, and augmented reality (AR) enhancements.

 GAI plays a crucial role in enhancing AIGC applications such as ChatGPT within xG wireless networks. By deploying GAI models at edge devices instead of relying solely on centralized cloud servers, network traffic and latency can be reduced. Additionally, GAI can create synthetic data for training AI models while preserving user privacy, unlocking entirely new possibilities for AIGC applications in xG networks. This approach has the potential to significantly change how AIGC is deployed in wireless networks \cite{AIGC-Survey}. 

\vspace{0.2cm}
\noindent
\textbf{Integrated sensing and communications (ISAC):} This technology is a key concept in 6G that aims to revolutionize wireless networks by breaking the traditional barrier between communication and sensing functionalities in wireless networks. Instead of operating as separate systems, ISAC allows the network infrastructure itself to act as a distributed sensor. 

ISAC exploits the unique properties of radio waves used for communication. By analyzing how these waves propagate, reflect, and scatter, the network can gather valuable information about its surroundings. Network elements like base stations (BSs), user devices, and even reconfigurable intelligent surfaces (RIS)\footnote{In RIS, a surface composed of numerous small, reconfigurable elements is deployed. This surface can adaptively manipulate the phase and/or amplitude of the incoming electromagnetic waves. By doing so, RIS can control the propagation of wireless signals, enabling beamforming, signal focusing, interference mitigation, and other advanced techniques to enhance communication performance.} become data collection points. The information gathered includes user location and mobility patterns, as well as signal propagation characteristics. This allows for dynamic resource allocation, optimizing network performance and user experience.

Leveraging GAI in ISAC can unlock its full potential. GAI-based frameworks not only provide comprehensive historical sensing data but also enable predicting future environmental changes. This capability adapts to variations in user behavior and environmental conditions, ensuring the broad applicability of GAI in ISAC systems \cite{GenISAC1}.

\vspace{0.2cm}
\noindent
\textbf{Semantic communications (SemCom):} SemCom focuses on conveying the meaning of the information being transmitted, rather than just the exact data bits. SemCom utilizes techniques from ML and information theory to understand the meaning of the data being transmitted. This could involve recognizing patterns, relationships, or even the intended purpose of the information. By exploiting inherent redundancy in data formats, SemCom compresses data packets by transmitting only the essential information needed to convey the meaning. It can take into account real-time network conditions such as channel quality or congestion. By understanding the content, the system can adjust transmission strategies such as power levels to prioritize the most critical information for successful communication. SemCom can improve network efficiency and reliability, reduce latency, and support innovative applications.

In SemCom, a key challenge lies in efficiently designing the encoders and decoders that handle semantic processing. Traditionally, both the encoder and decoder require joint training, making the process complex and energy-demanding. Recently, GAI has emerged as a potential solution to overcome these limitations, particularly in decoder design. GAI models excel at efficiently retrieving the original information (source information) without the need for joint encoder training. This capability makes GAI a promising tool for developing robust and efficient semantic decoders in SemCom systems \cite{GenSemCom1}.

\vspace{0.2cm}
\noindent
\textbf{Security of xG communications:} 
 Secure communication in xG networks goes beyond traditional notions of data encryption and access control. It encompasses protecting the confidentiality, integrity, and availability of data across the entire network infrastructure, from user devices to network core elements. This includes protecting against a wide range of threats, such as cyberattacks, eavesdropping, and unauthorized access. The dynamic nature of xG networks, characterized by features such as network slicing and integration of diverse technologies, necessitates adaptable security solutions capable of learning and evolving alongside potential threats. Accordingly, GAI would be a promising tool for enhancing security in xG communications.

\subsection{GAI for Network Optimization}
Based on what is discussed above, the power of GAI for xG network optimization lies in its two-fold approach: learning from vast datasets and generating new scenarios. GAI models are trained on real-world information critical for xG networks, including network traffic patterns, signal propagation characteristics, and resource allocation history, excelling at uncovering hidden patterns and relationships. This knowledge becomes the foundation for GAI's ability to generate new and realistic scenarios. A vast range of possibilities can be explored through these generated scenarios, which represent potential network conditions, eliminating the need for time-consuming and resource-intensive real-world data collection. The simulated scenarios can encompass diverse situations like fluctuations in user traffic, variations in signal propagation, or even entirely new network configurations. Optimal network configurations and strategies that maximize efficiency and user experience in xG networks can be identified by testing and evaluating network performance under these diverse GAI-generated scenarios. This approach also facilitates the development of robust configurations that function well in various real-world conditions. 

Based on their mentioned ability for learning and scenario generation, GAI offers several key applications for optimizing xG wireless networks. One crucial area is proactive resource allocation for efficiently utilizing bandwidth and infrastructure by adapting resources to anticipated demands and constraints. Another application lies in channel modeling. GAI can generate realistic channel conditions that encompass diverse propagation scenarios, overcoming the limitations of real-world xG network measurements. Accordingly, integrating GAI into xG networks can lead to a more efficient use of network resources, improved overall performance in xG networks, and accelerated exploration of novel technologies and network designs for xG networks.

However, deploying GAI models for networking applications presents unique challenges.  These challenges include the massive size of GAI models, requiring significant computational power and network bandwidth for transmission. Additionally, training effective GAI models for networking tasks often necessitates large and specialized datasets.  Fortunately, advancements such as distributed learning techniques, edge computing, on-device processing capabilities, and the utilization of GPUs for parallel computing, pave the way for efficient GAI deployment within networking environments. The synergy between GAI and xG networks is explored in the following. 

\subsection{Synergy of GAI and xG Networks}
The xG wireless networks will be characterized by high-dimensional configurations, non-linear relationships, and complex decision-making processes. As has been mentioned above, the ability of learning and scenario generation for the GAI models enable them to develop robust and efficient configuration for real-world applications in xG wireless networks. Therefore, recent research in xG wireless network optimization has focused on leveraging the transformative potential of GAI, marking a groundbreaking milestone extending AI's horizons beyond conventional boundaries. 

GAI can optimize mobile \emph{AIGC} by efficiently utilizing computational resources across the entire network, benefiting both users and providers. Additionally, through efficient user allocation and by selecting AIGC service providers with sufficient resources to manage content storage effectively, the network can be optimized in terms of bandwidth \cite{AIGC8}.

In \emph{SemCom}, GAI offers possibilities for making it more personalized, efficient, and reliable. GAI can enable context-aware communication, improve semantic reliability, and reduce misunderstandings \cite{GenSemCom4}. GAI-based optimization for covert communication aligns with SemCom applications requiring high levels of security and privacy \cite{GenSemCom5}. To grasp deeper meaning beyond explicit communication, employing GAI models allows for richer communication experiences \cite{GenSemCom7}. AI-generated contracts showcase GAI's potential for establishing trust and promoting efficient resource utilization within decentralized 6G networks with SemCom \cite{AIGC2}. Finally,  for optimal bandwidth allocation in SemCom-enabled networks, GAI models demonstrate the ability to handle complex network dynamics \cite{AIGC6}.

GAI has the potential to significantly enhance \emph{ISAC} systems, transforming how they capture, interpret, and transmit data. They offer significant potential for optimizing ISAC-enabled networks by providing accurate and comprehensive channel state information (CSI). They can improve signal processing by removing noise and reconstructing incomplete data, leading to more reliable communication \cite{GenISAC5, GenISAC6}.  GAI can also optimize network parameters for user association, and enable proactive handover decisions in order to minimize user disruptions on the move \cite{GenISAC1}.  

GAI models can offer a new frontier in \emph{securing xG communication} networks. Their ability to enrich the training dataset with a wide range of synthetic data that mimics real network behaviors can improve intrusion detection accuracy, model training efficiency, and data handling at the network's edge \cite{GenSec2}. They can enhance anomaly detection accuracy and handling of data imbalance as in \cite{GenSec4}. Data privacy and integrity, collaborative learning efficiency, adaptability and scalability of intrusion detection systems \cite{GenSec5}, and trust management \cite{GenSec10} can be improved by using the GAI models.

Several works in the literature survey the application of AI and GAI in enabling the above technologies within xG wireless networks. In the following, we briefly review these studies.

\subsection{Related Works}\label{RelWork}
 In the domain of {\em AIGC}, the study in \cite{AIGC10} provides an overview of ChatGPT, a powerful language model. It explains the foundation of ChatGPT, which is based on pre-trained LLMs trained on massive amounts of text data. These LLMs enable ChatGPT to understand and generate human-like language. Additionally, the study discusses how ChatGPT utilizes human feedback to improve its performance, allowing it to refine its responses and better align with human expectations.  It also highlights the potential risks associated with ChatGPT's misuse, such as the spread of misinformation or the automation of tasks that could displace human workers. 
Regarding the applications of AIGC, its core aspects such as general architecture, enabling technologies, working modes, key characteristics, and its potential through modern prototypes are explored in \cite{AIGC14}. This study discusses the challenges of security, privacy, and ethics in AIGC, analyzing potential threats and existing defense mechanisms. Furthermore, it explores intellectual property protection concerns and reviews techniques for safeguarding AIGC models and content which can lead  to the development of a more efficient, secure, and trustworthy AIGC ecosystem. 
In \cite{GenIoV1}, the challenges in enabling vehicular networks like limited bandwidth and communication latency are addressed by proposing a GAI-based framework. This framework is the one with a multi-modal architecture for handling both text and image data. 

Deep reinforcement learning (DRL) is used to optimize resource allocation and maximize the overall QoE within the network's limitations. The authors in \cite{AIGC15} study NetGPT, an approach for personalized LLMs. NetGPT leverages a collaborative cloud-edge architecture, deploying a distributed system with smaller, efficient LLMs positioned at the network edge (closer to users). This empowers NetGPT to personalize user experiences by utilizing location-based information and the edge LLM's ability to predict trends and infer user intent. Furthermore, by fostering deeper integration between communication and computing resources, NetGPT promotes a more AI-native network architecture. The collaborative approach of NetGPT, with its logical AI workflow, addresses resource limitations at the edge while enabling enhanced LLM synergy between edge and cloud. 

Regarding the ability of deep generative models (DGMs) to create realistic and complex data, the authors in \cite{AIGC16, AIGC-Survey1, AIGC-Survey} highlight the strengths and wide applicability of DGMs and propose  frameworks that incorporate DGMs into network optimization. Specifically, \cite{AIGC-Survey1} showcases practical applications through a series of case studies. These include integration with DRL for network control, designing incentive mechanisms to influence network behavior, and optimizing communication in IoV networks. The survey in \cite{AIGC-Survey} focuses on mobile AIGC networks, an approach that leverages AI to provide customized AIGC services at the network edge. By explaining generative models and the lifecycle of AIGC services within this mobile architecture and highlighting the collaborative cloud-edge-mobile infrastructure, the study explores the applications for text, image, video, and 3D content generation. While showcasing the potential benefits of mobile AIGC networks, the research discusses implementation challenges such as resource allocation, security concerns, and privacy risks.

Turning to SemCom in the xG wireless networks, \cite{GenSemCom3} investigates how to overcome limitations in context reasoning and background knowledge (BK). Recognizing the potential of GAI for creating diverse and personalized content, this work explores its application to SemCom. The proposed solution is a GAI-assisted SemCom Network (GAI-SCN) framework. This cloud-edge-mobile architecture leverages GAI models at both global and local levels, enabling capabilities like multimodal semantic content creation, semantic-level coding, and AI-generated content integration. However, challenges such as high computational demands and potential AIGC reliability issues remain to be addressed. The causal semantic communication (CSC) is studied in \cite{SemCom7} to tackle the challenge of bandwidth (BW) limitations in the context of digital twin (DT)-based wireless systems. Inspired by imitation learning, CSC leverages the DT's knowledge to train the receiver via SemCom over a BW-constrained channel. The transmitter acts as a teacher, identifying causal relationships within the data and transmitting causally-invariant semantic representations. The receiver, as the apprentice, employs a semantic decoder and builds a network state model to understand the environment dynamics. To overcome the limitations of imitation learning methods, CSC utilizes model-based reinforcement learning and semantic information metrics which is based on integrated information theory.

In the context of securing communications systems, \cite{GenSec6} surveys the current state-of-the-art research on leveraging GANs to address challenges within intrusion detection systems (IDS). This  review encompasses not only existing research on GAN-based IDS techniques, but also the specific datasets used for evaluation, the design methodologies of the employed GAN models, and the metrics utilized for performance assessment. By emphasizing on healthcare-specific privacy concerns, the study in \cite{GenSec19, GenSec21} explores the role of federated learning (FL) in preserving privacy within smart healthcare systems that utilize internet of medical things (IoMT) devices. The authors study how recent AI  models, such as DRL and GANs \cite{GenSec19}, and blockchain \cite{GenSec21}, can be integrated to enhance privacy protection within FL for IoMT networks. The survey in \cite{GenSec20} focuses on DL-based methods for digital watermarking and steganography. To address the issue of lacking a dedicated examination of deep watermarking, it explores DL techniques in both digital watermarking and steganography. It categorizes existing DL models based on their application (watermarking or steganography) and noise injection methods.  By supporting the idea of  the unification of watermarking and steganography under a software engineering approach, \cite{GenSec20} emphasizes the importance of building a more secure and trustworthy digital environment. {In \cite{GenSec23}, complex networks with a {\em $k$-core structure} are considered. The $k$-core structure of a network graph is defined as the largest subgraph in which each node has at least $k$ neighbors.} This study investigates the security threat through which malicious actors can disrupt the network by strategically removing edges, particularly targeting the highly connected innermost core. Two heuristic algorithms are proposed to efficiently identify critical edges for removal. The survey in \cite{AdvML} provides a systematic analysis of the impact of adversarial machine learning (AML) across all layers of wireless and mobile systems, from physical signals to network and application layers. It delves into the state-of-the-art techniques for generating and detecting adversarial samples, exploring methods such as GANs for crafting malicious data and the fast gradient sign method (FGSM) for identification. Additionally, it examines AML from both attacker and defender perspectives, emphasizing on methods for defense using adversarial models and reinforced learning.

\subsection{Motivation, Contributions, and Organization}
While existing research provides a valuable foundation for understanding AI and GAI in wireless networks, this paper takes a distinct approach.  Our focus is on GAI's transformative potential for optimizing xG wireless networks, with a particular emphasis on its application in 6G networks.  This comprehensive paper offers a structured learning experience, starting with the fundamentals of GAI models and progressing to practical applications through use cases. Recognizing the potential synergy between GAI and existing data-driven approaches commonly used in network optimization, this paper showcases the performance gains achievable by leveraging GAI's unique capabilities. This combination of theoretical knowledge and practical application bridges the gap between understanding GAI and its real-world implementation for optimizing xG networks, particularly focusing on the advancements it can bring to 6G technology.

Our contributions can be summarized as follows: 

\vspace{0.2cm}
\noindent
\textbf{GAI models for xG network optimization:} We review GAI models with the potential for optimizing xG networks, including generative flow networks (GFlowNet), generative adversarial networks (GANs), and generative diffusion models (GDMs). 

\vspace{0.2cm}
\noindent
\textbf{GAI for optimization and enhanced resource management of xG Networks:} 
We explore the use cases illustrating how these GAI models contribute to ongoing resource management and optimization in wireless networks, leading to ultimately enhanced overall network performance.

\vspace{0.2cm}
\noindent
\textbf{Integrating GAI with existing AI for xG network intelligence:} We introduce a new case study on utilizing GAI for resource allocation in non-terrestrial networks (NTNs) employing carrier aggregation (CA). This section delves into how GAI can improve the exploration in high-dimensional space and facilitates learning optimal resource allocation strategies, enhancing the intelligence and efficiency of xG networks.

The rest of the paper is organized as follows. The background of the GAI models is presented in Section~\ref{ModelGAI}.  Following this, Sections~\ref{AIGCOpt}, \ref{ISACOpt}, and \ref{SemComOpt} review case studies where GAI-based schemes are derived for optimization and resource allocation in the networks equipped with AIGC, ISAC, SemCom technologies, respectively. Employing GAI for security in xG communication is studied in \ref{xGSecurityOpt}.   Finally, Section~\ref{Challeng} highlights future studies and challenges for employing GAI models in xG wireless networks. The case study and conclusion are presented in Sections~\ref{NTN} and~\ref{Conc}, respectively.

\section{Basics of the GAI Models}\label{ModelGAI}
In this section, we delve into the mathematical foundations underpinning GAI models which are used by the studies in literature. Generative models, such as GFlowNets, GANs, and GDMs leverage statistical and algebraic concepts to generate new data instances that mimic the distribution of a given dataset. Let us denote the dataset by $\X = \{\mathbf{x}_i | i = 1, \ldots N\}$. In what follows, we will explore key mathematical principles including probability distributions, optimization algorithms, and loss functions, which facilitate the training and convergence of these models.

\subsection{Generative Flow Networks (GFlowNets)}
GFlowNets are a relatively new class of probabilistic models that have been designed to address the problem of efficient and unbiased sampling from complex, high-dimensional distributions \cite{GNetFlow}. Unlike other generative models such as GANs, GFlowNet focuses on learning to sample from distributions over combinatorial or structured spaces where the elements have varying sizes or complexities. {In xG wireless networks, GFlowNet training can be used to identify the optimal encoding and decoding methods in SemCom \cite{GenSemCom4}.} The details of GFlowNet fundamentals, and training objectives are given below.

\begin{figure}[t!]
  	\centering
  	\includegraphics[width=\columnwidth]{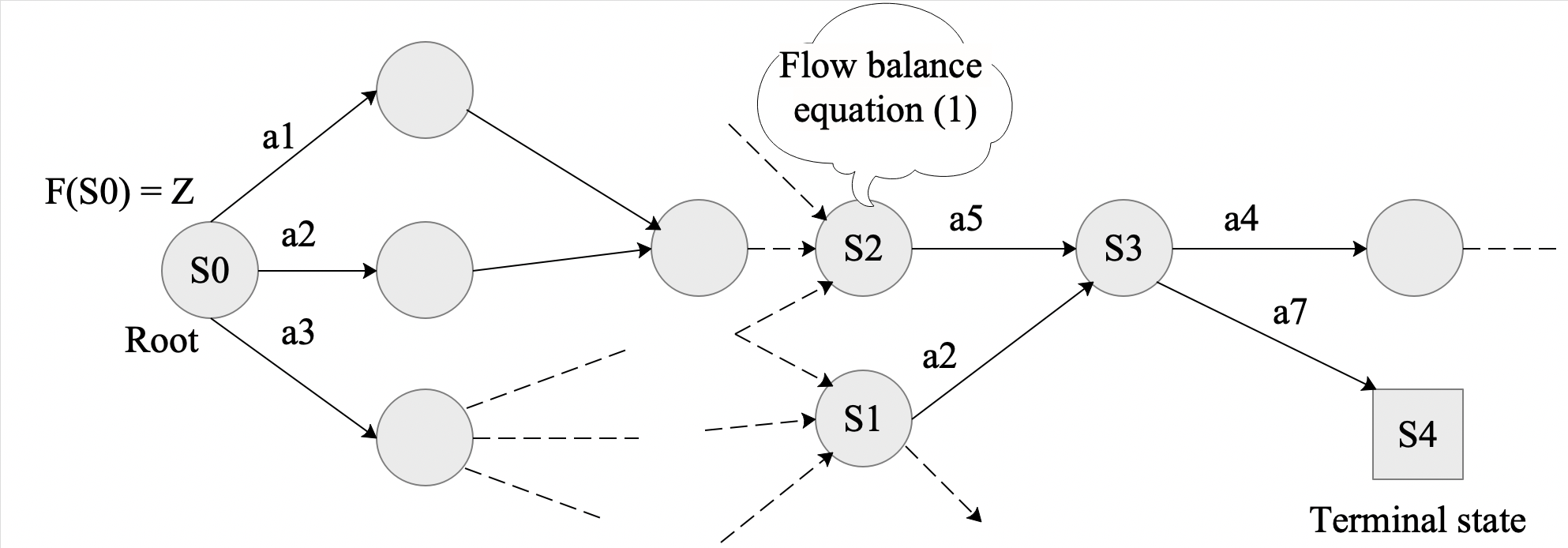}
  	\caption{Flow network ($F$) in a GFlowNet architecture: the root node ($S0$) starts with an inflow $Z$, and transitions between states ($S1, S2, S3$) occur via actions ($a1, a2, a3, a4, a5, a7$). Each state follows the flow balance equation ensuring the sum of incoming flows equals the sum of outgoing flows. The terminal state ($S4$) has an outflow denoted as $R(S4)$.}
  	\label{fig_GFlowNet}
  \end{figure}

As depicted in Fig.~\ref{fig_GFlowNet}, the structure of a Markov decision process (MDP) can be used for a flow network. Specifically, the flow network operates over a defined state space $S$, beginning from an initial state $s_0$ and progressing towards a set of terminal states $S_T \subset S$. Each state transition is associated with a flow $F(s, s^\prime)$, representing the probability of moving from state $s$  to state $s^\prime$. For each terminal state $s_T \in \s_T$ (e.g., $s_4$ in Fig.~\ref{fig_GFlowNet}), the reward function $R(s_T)$ is defined, reflecting their desirability. The rewards directly influence the flow values, i.e., higher rewards lead to higher flow into those states, guiding the network towards more desirable outcomes. The core principle governing GFlowNets is the balance of flow, where the flow into any state must equal the flow out, except at terminal states where the incoming flow should match the state's reward. In more detail, for a given not-terminal state $s$, the flow balance is given by \cite{GNetFlow}:
\begin{equation}\label{Flow_non_Terminal}
\sum\limits_{s_{\mathrm{pre}} \xrightarrow{} s} F(s_{\mathrm{pre}}, s) = \sum\limits_{s^\prime \in \mathrm{Children}(s)} F(s, s^\prime),
\end{equation}
where $s_{\mathrm{pre}}$ is a state before the state $s$ and the function $\mathrm{Children}(s)$ denotes the set of all possible states that can be reached directly from $s$. Additionally, the flow balance for a terminal state $s_T$ is given by \cite{GNetFlow}, 
\begin{equation}\label{Flow_Terminal}
F(s_T) = Z.R(s_T),
\end{equation}
where $Z$ is a normalization constant, essentially acting like a partition function in statistical mechanics.

The training objective of GFlowNets focuses on ensuring that the flows through the network accurately reflect the target distribution dictated by the rewards of terminal states. To achieve this, the networks are trained to minimize a loss function that quantifies the discrepancy between the actual flows and the ideal flows dictated by the reward function. The typical form of this loss function is given by \cite{GNetFlow},
\begin{equation}\label{GNetFlow_Loss}
L(\theta) = \sum\limits_{s \in S}\big(F_{\theta}(s) - \sum\limits_{s^\prime \in \mathrm{Children}(s)}F_{\theta}(s, s^\prime)\big)^2,
\end{equation}
where $F_{\theta}(s)$ represents the total flow into state $s$, parameterized by $\theta$, i.e., $F_{\theta}(s) = \sum\limits_{s_{\mathrm{pre}} \xrightarrow{} s}F_{\theta}(s_{\mathrm{pre}}, s)$, and $F_{\theta}(s, s^\prime)$ represents the flow from state $s$ to its child state $s^\prime$. This loss effectively encourages the network to balance the incoming and outgoing flows at each state, except at the terminal states where the incoming flow should be proportional to the reward. By minimizing this loss, GFlowNets learn to generate samples from a distribution that mirrors the desired probabilities, making them particularly powerful for tasks where the goal is to explore diverse high-reward solutions in complex environments. 

The sampling policy in GFlowNets is derived from the flows, where the probability of choosing an action from a given state is proportional to the flow associated with that action. For a given state $s$ and action $a \in \A(s)$ where $\A(s)$ denotes the set of feasible actions at state $s$, the sampling policy is given by \cite{GNetFlow}:
\begin{equation}\label{GNetFlow_Policy}
\pi(a|s) = \dfrac{F\big(s,T(s,a)\big)}{\sum\limits_{a^\prime \in \A(s)}F\big(s,T(s,a^\prime)\big)},
\end{equation}
where $T(s,a)$  denotes the state transition function. Based on the above discussion, the main steps involved in training GFlowNets to ensure the flow balance and optimization according to the specified loss function are given through the \textbf{Algorithm~\ref{alg_GNetFlow}}.
\begin{algorithm}[t!]
\SetKwFunction{Range}{range}
\SetKw{KwTo}{in}\SetKwFor{For}{for}{\string:}{}
\SetAlgoNoLine%
\textbf{Initialization}: Set parameters $\theta$ randomly;\\
\While{\text{not converged}}{
Initialize batch of states;\\
\For{\text{each state $s$ in batch \textbf{do}}}{
\If{\text{$s$ is terminal}}{
Compute reward $R(s)$;\\
Set incoming flow $F_{\theta}(s) = R(s)$;\\
}
\Else{
Compute the total incoming flow, i.e., $F_{\theta}(s) = \sum\limits_{s_{\mathrm{pre}} \xrightarrow{} s} F_{\theta}(s_{\mathrm{pre}}, s)$;
}
Compute total outgoing flow, i.e., $F_{\mathrm{out}}(s) = \sum\limits_{s^\prime \in \mathrm{Children}(s)}F_{\theta}(s,s^\prime)$;\\
\If{\text{$s$ is not terminal}}{
Update parameters $\theta$ to minimize the loss function, i.e., $\min \big(F_{\theta}(s) - F_{\mathrm{out}}(s)\big)^2$;
}
Select action $a$ by using \eqref{GNetFlow_Policy};\\
Execute action $a$ to transition to next state $s^\prime$;
}
Update parameters $\theta$ using gradient descent or another optimization method;
}
\caption{\textbf{GNetFlow} algorithm}
\label{alg_GNetFlow}
\end{algorithm}

\subsection{Generative Adversarial Network (GAN)} 
\begin{figure}[t!]
  	\centering
  	\includegraphics[width=\columnwidth]{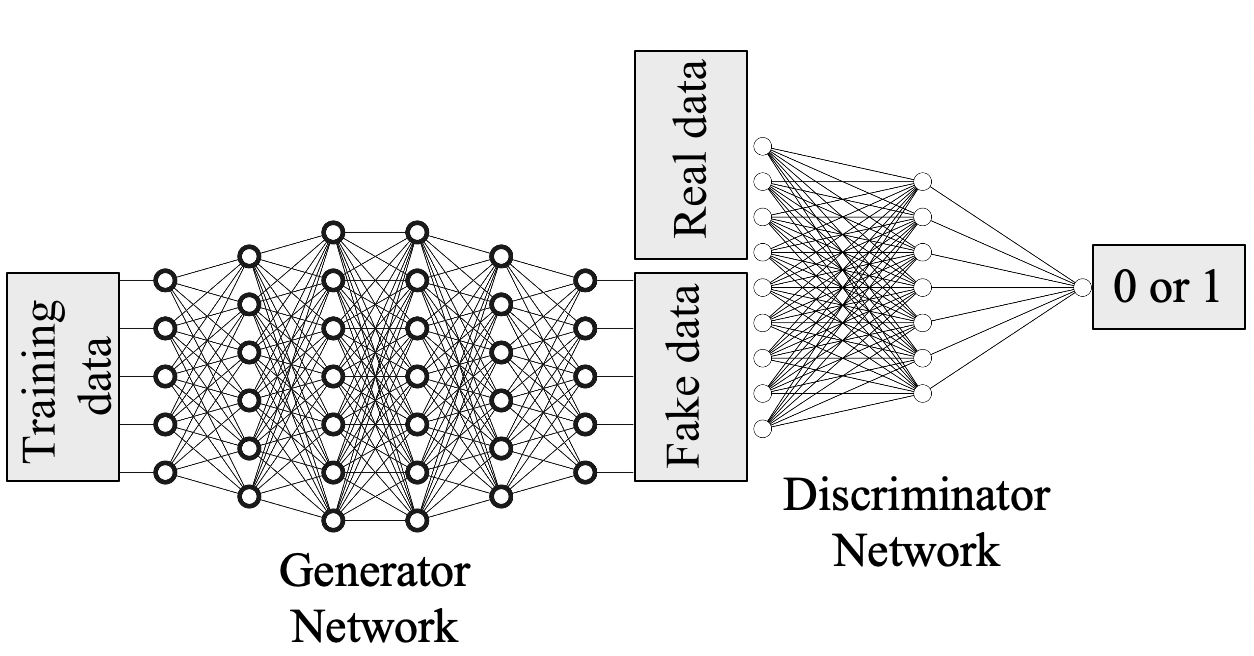}
  	\caption{Architecture of a GAN: the generator network takes training data and produces fake data, while the discriminator network distinguishes between real and fake data, outputting $0$ or $1$ depending on whether the input data is real or generated.}
  	\label{fig_GAN}
  \end{figure}
GANs are a powerful ML framework for learning probability distributions of various data types. As discussed in \cite{GAN} and illustrated in Fig.~\ref{fig_GAN}, this framework is based on an adversarial process where two multilayer perception models compete: a generator and a discriminator. 
During training, both models are trained simultaneously. The generator's objective is to create data samples that closely mimic a target distribution, learning to fool the discriminative model into making mistakes. While the discriminator strives to differentiate between true samples and the generator's creations. {In xG communications, GANs can be used for channel modeling and estimation, anomaly detection, and privacy preservation, as discussed in \cite{GenISAC3, GenISAC4, GenSec4}.} This framework corresponds to a minimax two-player game. The details for the game are given below.

Let $D_\omega(\mathbf{x}):\mathbb{R}^n \xrightarrow{} [0,1]$ and $G_\theta(\mathbf{z}): \mathbb{R}^d \xrightarrow{} \mathbb{R}^n$ denote the discriminator function and the generator function, respectively. The generator $G_\theta$ turns random samples $\mathbf{z} \in \mathbb{R}^d$ from distribution $\gamma$ into generated samples $G_\theta(\mathbf{z})$. In this setup, there are two main players: the discriminator $D_\omega$ and the generator $G_\theta$. The discriminator's role is to differentiate between real training samples and those generated by $G_\theta$. Essentially, it acts as a detective, attempting to spot the differences between genuine and fake data. On the other hand, the generator aims to produce samples that are as similar as possible to the real data, effectively trying to fool the discriminator into believing its creations are authentic.  Therefore, the objective of the game is to minimax the following target:
\begin{equation}\label{eq-GAN-value}
\begin{aligned}
& V(D_\omega,G_\theta) = E_{\mathbf{x} \sim \mu}[\log D_\omega(\mathbf{x})] + \\
& ~~~~~~~~E_{\mathbf{z} \sim \gamma}[\log (1 - D_\omega(G_\theta(\mathbf{z})))] 
\end{aligned}
\end{equation} 
In \eqref{eq-GAN-value}, the explicit expression for the target distribution $\mu$ is not available. However, $V(D_\omega,G_\theta)$ can be approximated by using the samples. Specifically, let $\M$ be a subset of samples from the training data set $\X$ and $\B$ be a minibatch of samples in $\mathbb{R}^d$ drawn from the distribution $\gamma$. The expectation values in \eqref{eq-GAN-value} can be approximated as follows:
\begin{subequations}\label{eq-approx}
\begin{align}
& E_{\mathbf{x} \sim \mu}[\log D_\omega(\mathbf{x})] \approx \dfrac{1}{\mid \M \mid} \sum\limits_{\mathbf{x} \in \M}\log D_\omega(\mathbf{x}),\\
& E_{\mathbf{z} \sim \gamma}[\log (1 - D_\omega(G_\theta(\mathbf{z})))] \approx \dfrac{1}{\mid \B \mid} \sum\limits_{\mathbf{z} \in \B}\log (1 - D_\omega(G_\theta(\mathbf{z}))).
\end{align}
\end{subequations}
Based on the above, the \textbf{GAN} algorithm is given in \textbf{Algorithm~\ref{alg_GAN}}.
\begin{algorithm}[t!]
\SetKwFunction{Range}{range}
\SetKw{KwTo}{in}\SetKwFor{For}{for}{\string:}{}
\SetAlgoNoLine%
    \For{\text{number of training iterations \textbf{\emph{do}}}}{ 
        \For{\text{$k$ steps \textbf{\emph{do}}}}{
        Samples minibatch $\{\mathbf{z}_i \mid i = 1, \ldots, m\}$ where $\mathbf{z}_i \in \mathbb{R}^d$ from distribution $\gamma$;\\
        Samples minibatch $\{\mathbf{x}_i \mid i = 1, \ldots, m\} \subset \X$ from the training set $\X$;\\
        Update the parameters for the discriminator $D_\omega$ by ascending its stochastic gradient with respect to $\omega$, i.e., $\nabla_\omega \dfrac{1}{m}\sum\limits^m_{i = 1} \big[\log D_\omega(\mathbf{x}_i)  + \log (1 - D_\omega(G_\theta(\mathbf{z}_i)))\big]$;
        }
        Samples minibatch $\{\mathbf{z}_i \mid i = 1, \ldots, m\}$ where $\mathbf{z}_i \in \mathbb{R}^d$ from distribution $\gamma$;\\
        Update the parameters for the generator $G_\theta$ by descending its stochastic gradient with respect to $\theta$, i.e., $\nabla_\theta \dfrac{1}{m}\sum\limits^m_{i = 1}  \log (1 - D_\omega\big(G_\theta(\mathbf{z}_i))\big)$;
        }
 \caption{\textbf{GAN} algorithm}
\label{alg_GAN}
\end{algorithm}
In this algorithm, the number of training steps for the discriminator, denoted by the parameter $k$, is a hyperparameter. 
In the original GAN paper \cite{GAN}, the authors opted for the least computationally expensive option, setting $k = 1$ for their experiments. While the convergence of the algorithm has been heuristically studied in \cite{GAN}, training GANs is known to be a delicate and unstable process. As a result, guaranteeing their convergence can be challenging. Additionally, trained generative models may suffer from a lack of diversity, meaning they might focus on generating samples that exhibit only a few recurring patterns instead of exploring the full data distribution. Recently, diffusion models have emerged as an alternative approach for data denoising. We will delve into the details of diffusion models in the following section.

\subsection{Generative Diffusion Model (GDM)}
GDM is considered as one of the state-of-the-art approaches in generative modeling and has shown impressive results in generating high-quality images. In the forward process of this model, noise is gradually added to the data until the signal becomes entirely Gaussian noise. In the reverse process, the original image is recovered by learning to reverse the diffusion process. {GDM-based method can be used for spectrum management \cite{AIGC6}, power control \cite{GenSemCom5, GenSemCom6, AIGC2}, and load balancing at edge \cite{AIGC8} in xG communication networks.}

The denoising diffusion probabilistic model (DDPM) is a particular type of GDM that focuses on a specific way of defining the denoising steps. DDPMs explicitly model the transition from noisy data to clean data in a probabilistic manner. They learn how to remove a specific amount of noise from the data at each step, eventually converging on the original clean data distribution.
\begin{figure}[t!]
 	\centering
 	\includegraphics[width=\columnwidth]{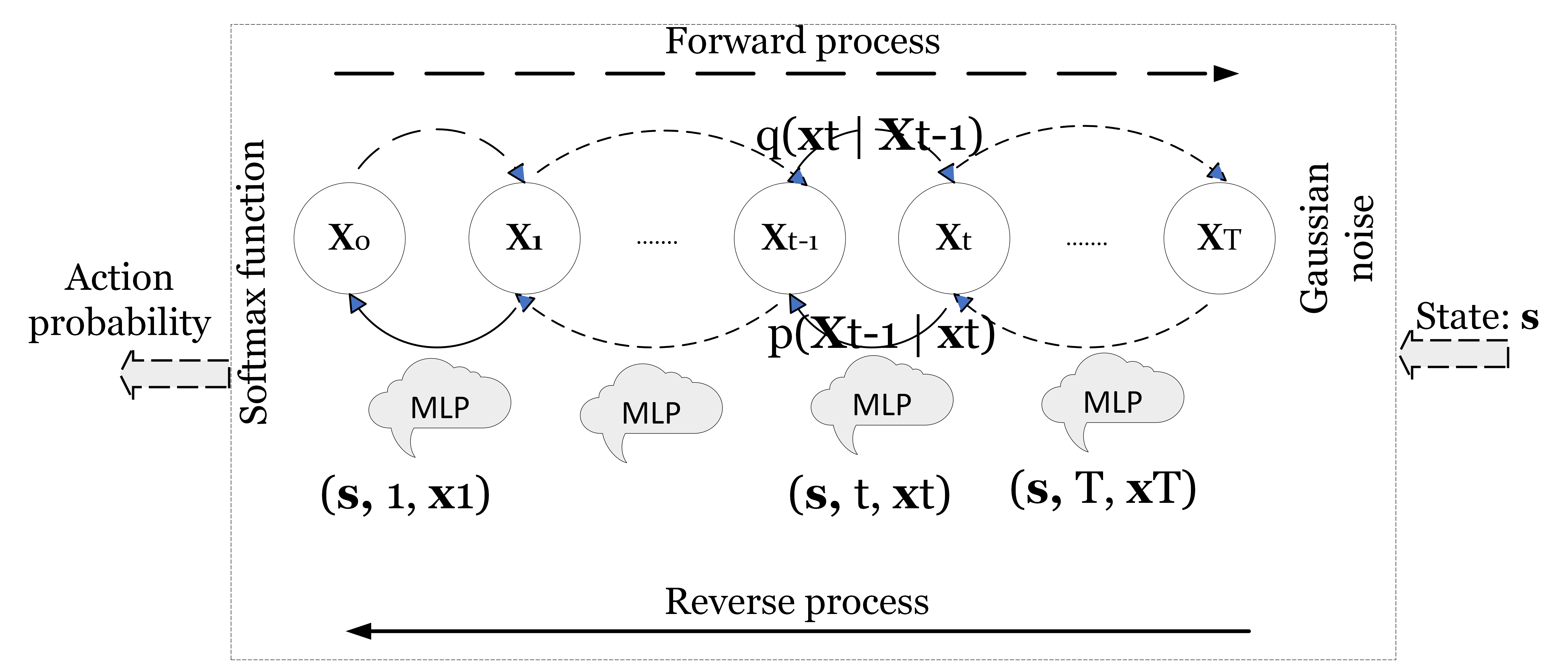}
 	\caption{Deriving the action probability given in \eqref{eq_softmax} through the backward process of diffusion-based GADM algorithm.}
 	\label{AGOD}
 \end{figure}
As illustrated in Fig.~\ref{AGOD}, in the forward process of DDPM, given a target probability distribution $\mathbf{x}_0$, a sequence of Gaussian noises is added to it to obtain $\mathbf{x}_1, \mathbf{x}_2, \ldots, \mathbf{x}_T$. The transition from $\mathbf{x}_{\tPrm-1}$ to $\mathbf{x}_{\tPrm}$ is defined as a normal distribution with mean $\sqrt{1-\beta_{\tPrm}}\mathbf{x}_{\tPrm-1}$ and variance $\beta_{\tPrm}\mathbf{I}$ as $q(\mathbf{x}_{\tPrm} \mid \mathbf{x}_{\tPrm-1}) = N(\mathbf{x}_{\tPrm}, \sqrt{1-\beta_{\tPrm}}\mathbf{x}_{\tPrm-1}, \beta_{\tPrm}\mathbf{I})$. It can be seen that $\mathbf{x}_{\tPrm}$ depends only on $\mathbf{x}_{\tPrm-1}$. Therefore, the forward process can be considered a Markov process and thus given $\mathbf{x}_0$, the distribution $\mathbf{x}_T$ can be given by $q(\mathbf{x}_T \mid \mathbf{x}_0) = \prod\limits_{t=1}^{T} q(\mathbf{x}_{\tPrm} \mid \mathbf{x}_{\tPrm-1})$. The forward process is used to establish the mathematical relationship between the target probability distribution $\mathbf{x}_0$ and the Gaussian noise image $\mathbf{x}_T$ as:
\begin{equation}\label{eq_xt_x0}
\mathbf{x}_{\tPrm} = \sqrt{\overline{\alpha}_{\tPrm}}\mathbf{x}_0 + \sqrt{1 - \overline{\alpha}_{\tPrm}}\Boldeps.
\end{equation}
In \eqref{eq_xt_x0}, $\Boldeps$ follows the standard Normal distribution, i.e., $N(\mathbf{0},\mathbf{I})$. The detail for hyper-parameter $\overline{\alpha}_{\tPrm}$ is given in \cite{DDPM}. In wireless network optimization, the lack of optimal decision solutions results in the forward process not being executed. Instead, it is utilized to derive equations in the reverse process, as explained in the following.    

In the reverse process, known as the sampling process, the noises are gradually removed to obtain the sequence $\mathbf{x}_{T-1}, \ldots, \mathbf{x}_0$. In other words, $\mathbf{x}_0$ is inferred from a noise sample $\mathbf{x}_T \sim N(\mathbf{0},\mathbf{I})$ by gradually removing the noise. In the backward Markov process in Fig.~\ref{AGOD}, the transition from $\mathbf{x}_{\tPrm}$ to $\mathbf{x}_{\tPrm-1}$ is defined by a normal distribution with mean $\Boldmu_{\Boldtheta}(\mathbf{x}_{\tPrm}, \tPrm, \mathbf{s})$ and variance $\Tilde{\beta}_{\tPrm}\mathbf{I}$, expressed as $p_{\Boldtheta}(\mathbf{x}_{\tPrm-1} \mid \mathbf{x}_{\tPrm}) = N\big(\mathbf{x}_{\tPrm-1}, \Boldmu_{\Boldtheta} (\mathbf{x}_{\tPrm}, \tPrm, \mathbf{s}), \Tilde{\beta}_{\tPrm}\mathbf{I}\big)$. Further details on the hyper-parameter $\Tilde{\beta}_{\tPrm}$ can be found in \cite{DDPM}. Moreover, $\Boldmu_{\Boldtheta}(\mathbf{x}_{\tPrm}, \tPrm, \mathbf{s})$ can be learned by a deep model, with specifics provided in what follows.

By applying the Bayesian formula, $\Boldmu_{\Boldtheta}(\mathbf{x}_{\tPrm}, \tPrm, \mathbf{s})$ can be formally expressed as a weighted sum of $\mathbf{x}_{\tPrm}$ and $\mathbf{x}_0$ \cite{DDPM}. Utilizing the equation from the forward process given in \eqref{eq_xt_x0}, $\Boldmu_{\Boldtheta}(\mathbf{x}_{\tPrm}, \tPrm, \mathbf{s})$ is finally expressed as $\Boldmu_{\Boldtheta}(\mathbf{x}_{\tPrm}, \tPrm, \mathbf{s}) = f\big(\mathbf{x}_{\tPrm}, \Boldeps_{\Boldtheta}(\mathbf{x}_{\tPrm}, \tPrm, \mathbf{s})\big)$, where $\Boldeps_{\Boldtheta}(\mathbf{x}_{\tPrm}, \tPrm, \mathbf{s})$ is a deep model parameterized by $\Boldtheta$ that generates denoising noises conditioned on the observation state $\mathbf{s}$ and $\tPrm = 1, \ldots, T$; additionally, the details for function $f(.)$ are given in \cite{DDPM}. Now, for reverse sequence $\mathbf{x}_T, \ldots, \mathbf{x}_0$, $\mathbf{x}_{t-1}$ can be sampled from reverse transition distribution $p(\mathbf{x}_{\tPrm})p_{\Boldtheta}(\mathbf{x}_{\tPrm-1} \mid \mathbf{x}_{\tPrm})$. This can be used to ultimately obtain the generation distribution $p_{\Boldtheta}(\mathbf{x}_0)$ expressed as $p_{\Boldtheta}(\mathbf{x}_0) =  p(\mathbf{x}_T) \prod\limits_{\tPrm=1}^{T} p_{\Boldtheta}(\mathbf{x}_{\tPrm-1} \mid \mathbf{x}_{\tPrm})$, where $p(\mathbf{x}_T)$ is a standard Gaussian distribution. Once the generation distribution $p_{\Boldtheta}(\mathbf{x}_0)$ is obtained, we
can sample the output $\mathbf{x}_0$ from it. 

One common challenge encountered when obtaining $p_{\Boldtheta}(\mathbf{x}_0)$ using the above equation is the difficulty in adjusting the model's parameters due to the inherent randomness in sampling from a probability distribution, which can render traditional backpropagation ineffective. To address this challenge, parameterization can be utilized to separate the randomness from the distribution's parameters. Consequently, the following update rule is used \cite{DDPM}:
\begin{equation}\label{eq_update_rule}
\mathbf{x}_{\tPrm-1} = \Boldmu_{\Boldtheta}(\mathbf{x}_{\tPrm}, \tPrm, \mathbf{s}) + \big(\dfrac{\Tilde{\beta}_{\tPrm}}{2}\big)^2 \mathbf{.} \Boldeps,~\forall \tPrm = 1, \ldots T.  
\end{equation}
By iteratively applying the updating rule in \eqref{eq_update_rule}, all $\mathbf{x}_{\tPrm}, \tPrm = 1, \ldots, T$ and in particular $\mathbf{x}_0$ is obtained form a randomly generated normal noise $\mathbf{x}_T$. By applying the softmax function, $\mathbf{x}_0$ can be converted into a probability distribution, denoted as $\pi_{\Boldtheta}\big(\mathbf{s}\big)$:    
\begin{equation}\label{eq_softmax}
 \pi_{\Boldtheta}\big(\mathbf{s}\big) = \left\{ \dfrac{\exp{x_{0j}}}{\sum^{\mid \mathbf{x}_0 \mid}_{k=1}\exp{x_{0k}}}  \right\},~\forall j = 1, \ldots, \mid \mathbf{x}_0 \mid.
 \end{equation}
 
Consequently, the generative AI-based decision-making (GADM) algorithm is derived as in \textbf{Algorithm~\ref{alg_GADM}}. 
\begin{algorithm}[t!]
\SetKwFunction{Range}{range}
\SetKw{KwTo}{in}\SetKwFor{For}{for}{\string:}{}
\SetAlgoNoLine%
\SetKwInOut{Input}{input}
\SetKwInOut{Output}{output}
   \Input{State $\mathbf{s}$, and DNN parameter $\Boldtheta$;}
   \Output{Probability distribution $\pi_{\Boldtheta}\big(\mathbf{s}\big)$;}
   \textbf{Initialization}:
    $\mathbf{x}_T \sim N(\mathbf{0},\mathbf{I})$ and $\Boldeps \sim N(\mathbf{0},\mathbf{I})$;\\
    
    \For{$\tPrm = T: 1$ \text{\textbf{\emph{do}}}}{
        Calculate $\Boldeps_{{\Boldtheta}}(\mathbf{x}_{\tPrm}, \tPrm, \mathbf{s})$ from a DNN parametrized by $\Boldtheta$ and scale it as $\tanh\big(\Boldeps_{\Boldtheta}(\mathbf{x}_{\tPrm}, \tPrm, \mathbf{s})\big)$;\\   
        Obtain $\Boldmu_{\Boldtheta} = f\Big(\mathbf{x}_{\tPrm}, \tanh\big(\Boldeps_{\Boldtheta}(\mathbf{x}_{\tPrm}, \tPrm, \mathbf{s})\big)\Big)$;\\ 
        Calculate $\mathbf{x}_{\tPrm - 1}$ from equation \eqref{eq_update_rule};\\  
        }
        \textbf{end}\\
        Calculate the probability distribution for $\mathbf{x}_0$, i.e., $\pi_{\Boldtheta}\big(\mathbf{s}\big)$ by using \eqref{eq_softmax};\\ 
 \caption{\textbf{GADM} algorithm}
\label{alg_GADM}
\end{algorithm}
The \textbf{GADM} algorithm can be integrated into different DRL architectures. Specifically, the gathered experience (i.e., states, actions, rewards, and next states) in the replay buffer, which is a common component in many DRL algorithms, can be used to train the DDPM model and obtain the optimal $\Boldtheta$.

\subsection{Complexity of GAI Models}\label{Complexity}
While GAI models offer significant potential for optimizing various aspects of xG wireless networks, their practical deployment depends on managing their computational complexity. Different GAI models inherently possess varying complexity levels. For instance, GAN-based channel estimation complexity, as detailed in \cite{GenISAC5, GenISAC6}, is directly tied to the model architecture, including the number of convolutional layers, features per layer, mini-batch size, and iterations. Therefore, an approach analyzing factors such as model architecture, training processes, and inference requirements is required.

Research in \cite{GenSemCom6} explores the use of efficient layer aggregation network (ELAN) modules to reduce model parameters and computational complexity in object detection tasks. While originally developed for image recognition and computer vision (CV), ELAN modules hold promise for network optimization as well. Similarly, the GAI-based SemCom approach in \cite{GenSemCom5} achieves complexity reduction by eliminating the need for joint decoder-encoder training. Additionally, designing the interest point encoder within the semantic encoder with an explicit decoder, as suggested in \cite{AIGC2}, can further reduce computational cost. Here, the interest point encoder acts as a pre-processing step, identifying key locations (interest points) in the input data. This allows the semantic encoder to build a higher-level understanding using these identified points, leading to a more efficient model. Finally, feature selection techniques, as employed in \cite{GenSec2, GenSec5, GenSec10}, can also be leveraged to improve training speed and reduce complexity by focusing on the most relevant data features.

The following sections explore key communication paradigms in xG wireless networks, including SemCom, mobile AIGC networks, ISAC, and xG communication security. Integrating these technologies offers significant benefits for wireless networks, such as  enhanced safety, a richer user experience, and improved sustainability. We will then delve into how GAI can further optimize networks empowered by these technologies, and how it can address potential challenges associated with their implementation in xG communication.

\section{ GAI-Based Optimization in Mobile AIGC Networks}\label{AIGCOpt}

\subsection{Mobile AIGC Networks}

Mobile AIGC networks combine the power of AIGC with the low-latency benefits of mobile edge computing, allowing for personalized AIGC services delivered directly to mobile devices.  This functionality is achieved through seamless cooperation among computing and storage resources distributed across the cloud, the network edge, and even mobile devices themselves.

GAI models offer significant benefits in optimizing mobile AIGC networks. Throughout the various stages of the AIGC service lifecycle, pre-training, fine-tuning, and inference, GAI models can unlock efficiency gains by leveraging data generated locally on devices like IoT sensors and smartphones \cite{AIGC-Survey}.
Exceling in both reconfigurability and accuracy, GAI models can adapt seamlessly to evolving network demands and user preferences. This ultimately facilitates the production of a vast amount of personalized content \cite{AIGC-Survey}. Furthermore, deploying GAI models at the network edge offers additional sustainability benefits. By reducing reliance on large-scale, centralized storage infrastructure, GAI models can significantly decrease energy consumption and the corresponding carbon footprint associated with traditional AIGC operations. We begin this section by exploring how the GAI model is employed to optimize content generation within the Metaverse. We will then discuss the efficiency gains that can be obtained through this optimization process. 

While AIGC models offer vast potential for content generation in the Metaverse, their deployment can be hindered by the challenge of demanding training requirements. To address this, recent research \cite{AIGC8} proposes an AIGC-as-a-Service (AaaS) architecture where AIGC models are deployed at the network's edge. This strategic placement enables users to access these services ubiquitously, from any device and at any location. Motivated by this paradigm shift, the study in \cite{AIGC8} focuses on the problem of AIGC service provider (ASP) selection. The objective is to maximize user utility while considering the total resource availability constraints of each service provider. User utility cannot be fully known in advance and is instead determined dynamically using a human-aware content quality assessment function. This function might consider factors such as contrast, sharpness, and texture for image-based content. This service provider selection problem can be modeled as a RL system. To address this challenge, a deep diffusion RL-based method is employed, which leverages the combined strengths of both actor-critic model and DDPM. The following describes the specifics of this approach. 

Within the developed RL system, the state-space encompasses two key elements: total available resources across the network and the current resource availability of each ASP. The action space defines the set of all possible decisions for assigning the current Metaverse user task to a specific ASP. Finally, the reward function is designed in terms of the user utility. To achieve optimal service provider selection, this RL system aims to train the parameters for the DDPM model. This is accomplished through an actor-critic architecture consisting of actor networks (target and online), critic networks (target and online), and replay buffer. The core of the actor-network is based on the diffusion model illustrated in Fig.~\ref{AGOD}. The parameters for the DNNs in actor-network are updated to maximize the expectation of Q-values\footnote{The Q-values estimate the expected cumulative reward for each action at the current state.} over all actions to improve the policy for choosing ASPs. Meanwhile, the critic network, specifically a double-critic network, reduces overestimation bias and evaluates ASPs' selection. During training, the Q-value (which is used to update the actor-network) is the minimum of the two Q-value estimates from the two critic networks. The critic networks, on the other hand, are trained to minimize the temporal difference (TD) error between the target Q-value (i.e., the output of the target critic network) and the evaluated Q-value  (i.e., the output of the online critic network). Experiences from user interactions are stored in a replay buffer and used to update both the actor and critic networks.    

\subsection{How GAI Models are Used for Network Optimization}
In \cite{AIGC8}, the generative AI model (DDPM) plays a critical role in encouraging exploration within the selection process.  The DDPM's output effectively captures the dependencies between the user's observation space and the available ASPs (action space). This allows the RL agent to understand how different resource allocations at each ASP might impact its utility. By feeding the DDPM's output into a softmax function, the system generates a probability distribution for each potential ASP selection. This probability distribution, informed by the DDPM's predictions, guides the RL agent's policy function, encouraging it to not only prioritize high-utility selections but also explore less-explored options. This exploration aspect is crucial for the agent to continually learn and adapt to potential variations in resource availability and AIGC model performance over time.

Based on the above discussion, in \cite{AIGC8}, by guiding the RL agent's policy function through a diffusion-based approach, the system achieves efficient utilization of \emph{computational resources} across the entire network. This benefits both users (improved service quality) and providers (avoided system overload), ultimately maintaining network performance. Additionally, this GAI-driven approach indirectly contributes to network optimization in terms of \emph{bandwidth} (through efficient user allocation) and \emph{storage} (by selecting providers with sufficient resources to manage content storage effectively).

\section{GAI-Based Optimization in ISAC-Enabled Networks}\label{ISACOpt}

\subsection{GAI and ISAC-Enabled Networks}

ISAC is a design approach that merges sensing and communication functionalities. It leverages existing wireless communication systems for sensing purposes. The insights gained from this wireless sensing are then used to optimize the communication itself \cite{GenISAC2}. GAI can significantly enhance ISAC performance in wireless networks. In  ISAC systems, these models find applications in various areas, including data augmentation, anomaly detection, and time-series forecasting \cite{Survey-6G}. In the following, we will delve into the details of how GAI empowers ISAC.

GAI's data modeling and analysis capabilities hold promise for optimizing wireless communications systems with ISAC \cite{GenISAC3, GenISAC4}. GAI can improve \emph{channel estimation}, a critical component of ISAC, by generating synthetic data for training ML models. This addresses the challenge of limited real-world data typically available for channel estimation in ISAC systems.  Another key challenge for ISAC is efficient resource allocation. Here, GAI models can also be leveraged. By generating synthetic data for training ML models, GAI can optimize resource allocation, further enhancing the overall efficiency of ISAC systems in wireless networks.

\subsubsection{Application of GANs for channel estimation in RIS-aided communication systems}
The studies in \cite{GenISAC5, GenISAC6} investigate the application of GANs for channel estimation in RIS-aided communication systems. In \cite{GenISAC5}, the convolutional blind denoising GAN (GAN-CBD) \cite{GANCBD} is used to denoise channel data and improve CSI estimation. GAN-CBD offers a novel approach to image denoising. Unlike traditional methods requiring clean image examples for training, GAN-CBD leverages the power of GANs, effectively removing noise from images without prior knowledge of the specific noise type through the adversarial training process. It helps train the generator network to create more realistic outputs. The details are given in what follows. 

 In \cite{GenISAC5}, the generator network for the GAN-CBD system employs two subnetworks: a noise level estimation subnetwork and a non-blind denoising subnetwork. The noise level estimation subnetwork tackles the challenge of noise in the wireless channel. It receives the received signal, separated into its real and imaginary parts, and combines them for processing. Convolutional layers then analyze the combined signal to estimate the level of noise present.  Additional SoftMax layers interpret the results, providing a final, single-value estimate of the noise level. The non-blind denoising subnetwork focuses on cleaning the received signal and estimating the actual data channel. It leverages two inputs, the received signal (containing both data and noise) and the estimated noise level obtained from the previous subnetwork.  The combined signal and noise information are fed through residual block layers\footnote{Residual block layers are a key architectural element in CNNs. They address the vanishing gradient problem, a challenge where information weakens as it travels through many layers. These blocks introduce shortcut connections that allow the original input to be directly added to the processed output within the layer. This ensures the network retains important information even if processing weakens it.  This improvement in learning allows deep networks to train faster, achieve better accuracy, and ultimately perform more effectively on various tasks.}, which act like filters. These layers analyze both the signal and the estimated noise level to remove noise and recover the original, clean signal. Finally, after noise removal, the system estimates the actual channel the signal traveled through.  A loss function continuously monitors the subnetwork's performance by comparing the estimated noise level with the actual noise level. This comparison helps the system refine its noise removal and channel estimation capabilities over time.


The study in \cite{GenISAC6} proposes a novel approach to address the challenges of instability and inefficiency associated with using GANs for CSI estimation in intelligent reflecting surface (IRS) aided communication systems. The solution leverages model-driven DL techniques \cite{MD-DL} by incorporating prior knowledge about the inherent properties of the channels into the network structure of the GAN model. This approach facilitates faster training of the neural networks (NNs). In the scenario with one user, the generative model within the proposed IRS-GAN framework is comprised of three distinct nodes: a BS-IRS (BI) node, an IRS-User (IU) node, and a Cascading (C) node.
The BI node serves a crucial role in transforming an input random noise vector, initially drawn from a fixed distribution like uniform or Gaussian, into a matrix (denoted by $\Tilde{\mathbf{G}}$) that approximates the actual BS-IRS channel (denoted by $\mathbf{G}$). This transformation leverages the concept of treating the channel matrix $\mathbf{G}$ as a two-dimensional image. Consequently, a two-dimensional convolutional NN (CNN) is employed within the BI node to effectively capture the inherent correlations between the various elements within the channel matrix.
Many DL frameworks operate exclusively with real numbers. To address this limitation and handle complex-valued input channels like the channel matrix $\mathbf{G}$, the proposed approach decomposes $\mathbf{G}$ into its real and imaginary parts. These parts are then treated as separate image color channels with real-valued entries. Additionally, the Line-of-Sight (LoS) component within $\mathbf{G}$, which exhibits minimal temporal variation, is incorporated as a separate bias term independent of the input noise vector. 
Similar to the BI node, the IU node also transforms a random noise vector. The IU node employs a multi-layer fully-connected network for this transformation. Additionally, a bias term is added to account for any constant value present in the IRS-user channel.
The Cascading (C) node combines the information extracted by the BI and IU nodes. It first performs a concatenation operation on the outputs from these nodes, essentially merging them. This combined data is then passed through a series of fully-connected layers. These layers further refine the approximation of the reflected channel. Finally, the C node outputs the generated reflected channel samples. 
The study leverages the Wasserstein distance the generated channel distributions and the true channel distributions (\cite{WesGAN}) as the loss function for training the IRS-GAN framework. The Wasserstein distance is both continuous and differentiable everywhere. This property contributes to the improved training stability and convergence of the proposed IRS-GAN. The training process itself is iterative and consists of two main loops. Within each outer loop, three inner loops are executed sequentially: a generative loop, a discriminative loop, and a testing loop. During each inner loop, the corresponding network (generative or discriminative) undergoes training for a fixed number of iterations. The testing loop, on the other hand, focuses on estimating the Wasserstein distance between the true channel distribution and the generated channel distribution. This estimated distance is then stored for comparison in subsequent testing loops.

The proposed IRS-GAN framework demonstrates potential for extension to multi-user scenarios. This is because each reflecting element within the IRS reflects signals from the BS to various users through the same BS-IRS channel. Consequently, the reflected channel for any user can be considered a scaled version of the channels experienced by other users. By leveraging this insight, a single GAN can be constructed to learn the reflected channels for all users simultaneously. This approach offers a significant advantage compared to employing multiple independent, identical IRS-GANs. It allows for a simplified network structure and a reduction in the overall number of network parameters required. 
The proposed IRS-GAN framework scales efficiently to multi-user scenarios. It comprises a single generative model and $K$ independent discriminative models, where $K$ represents the number of users. The generative model leverages a single BI node to capture the strong correlation among the reflected channels inherently within its network structure. It further incorporates $K$ separate IU nodes and $K$ separate C nodes, each dedicated to a specific user. The discriminative models, meanwhile, maintain the same structure as in the single-user case. A key advantage of this proposed framework lies in the simultaneous training of the generative and discriminative models. This enables the learning of reflected channels for all users within a single training process. The loss function for the generative model incorporates the sum of Wasserstein distances between the generated channel distributions and the true channel distributions for each user. The loss function for each discriminative model focuses on a gradient penalty term. To accelerate training further, all $K$ discriminative models can be trained in parallel.

\subsubsection{Optimizing XR experiences over THz wireless systems with RIS}
The study in \cite{GenISAC1} tackles optimizing extended reality (XR) experiences over terahertz (THz) wireless systems using RIS. It achieves this by minimizing handover costs while ensuring quality of personal experience (QoPE) for XR users across the reality-virtuality spectrum, leveraging comprehensive and predictive sensing information.

As shown in \cite{GenISAC1}, by concatenating and tensor-factoring received signals from uplink (UL) snapshots, an optimization problem can estimate RIS subarray sensing parameters for user path attenuation factors. 
However, this estimation has limitations. It doesn't account for potential blockages, whether a line-of-sight (LoS) connection exists, or the user's degrees of freedom (DoF) during non-line-of-sight (NLoS) scenarios. Consequently, it fails to capture the complete picture of \emph{user behavior and environmental dynamics} in THz.
To address this, the study proposes a DL approach based on non-autoregressive (NAR) models through which the entire input is analyzed together to predict the entire output sequence simultaneously. This approach predicts the missing values within the LoS and NLoS sensing matrices, resulting in comprehensive and continuous sensing information. These comprehensive, time-varying vectors are then fed into an encoder-decoder based GAI framework for the prediction of future time slots based on continuous and comprehensive sensing data. The details are given in what follows.

The DL model uses a two-part architecture: forward-backward encoder for imputation, and multi-resolution decoder for forecasting. In \emph{forward-backward encoder}, the encoder processes incomplete sequences of LoS and NLoS sensing data. It utilizes a combination of forward recurrent neural network (FRNN) and backward recurrent neural networks (BRNN)\footnote{Information in an FRNN flows strictly forward through the network, one step at a time. FRNN struggles with long sequences because the information from earlier parts of the sequence can fade or disappear as it travels through the network. BRNNs address this by incorporating a backward pass of information, allowing the network to consider information from both the past and the future of a specific point in the sequence.} to model the conditional distribution of hidden states based on the available sensing information and masked inputs. In \emph{multi-resolution decoder}, the decoder leverages the hidden representations generated by the encoder to predict the missing sensing values. It can handle different time scales within the data, allowing the model to capture both short-term and long-term environmental changes. This multi-scale learning capability improves the generalizability and accuracy of the imputation process.

To proactively anticipate and respond to user and environmental changes, the system incorporates an \emph{encoder-decoder based transformer structure} with an \emph{auxiliary discriminator}. Encoder-decoder transformers excel at time series forecasting thanks to their multi-head attention layers. Multi-head attention layers within the encoder allow the encoder-decoder model to focus on the most relevant parts of the input sequence, attending to specific elements based on their importance to the overall meaning. This refined understanding is then passed to the decoder, which generates the output sequence based on the encoded information. Therefore, these layers allow the transformer to identify long-term dependencies within the continuous stream of imputed sensing data.
To understand relationships between elements within a sequence, a novel architecture called the $\alpha-$Exam transformer \cite{alphaExam} is chosen.  Fenchel-Young loss is used in this transformer which is a convex loss function for a regularized prediction \cite{alphaExam}. Additionally, to enhance the generalizability of the AI framework and effectively capture the inherent randomness (stochasticity) of the data, an adversarial training process is employed. This process improves the robustness of the model by regularizing the sensing data, enabling it to handle data variations and learn from both observed and unobserved information. This, in turn, strengthens the model's ability to generalize to new, unseen data.
Furthermore, a discriminator network is added to improve the accuracy of predictions across different time scales. This network consists of three fully connected linear layers with Leaky ReLU activation functions.

Leveraging comprehensive and predictive sensing information, the system defines the quality of personal experience (QoPE) for XR users across the reality-virtuality spectrum. To minimize disruptions caused by frequent handovers, a user-centric handover cost is also defined. This cost considers factors like the user's association with a specific RIS subarray, their movement speed, total travel distance, and the handover delay itself. By minimizing the overall handover cost, the system aims to maximize the collective utility for all active RIS subarrays. This optimization must be achieved while ensuring acceptable QoPE for users.
To address this challenge, the problem is first modeled as a multi-agent RL system. This model is then tackled using a hysteretic deep recurrent Q-network (HDRQN) approach \cite{rQ-net}. HDRQN learns complex patterns within the sequence of observations and actions which can be particularly helpful in tasks with long-term dependencies. In such tasks, the optimal action at any given point depends on a series of past events.

\subsection{How GAI Models are Used for ISAC-Enabled Network Optimization?}
\begin{table*}[t!]
\caption{Optimizing ISAC-enabled networks with GAI}
\label{tab_ISAC}
\begin{tabular}{|c||c||c||c||c|}
\hline
Ref.     & GAI model                                                                                                  & \begin{tabular}[c]{@{}c@{}}Optimizing\\ variable\end{tabular} & \begin{tabular}[c]{@{}c@{}}Objective/loss \\ function\end{tabular}                                                  & \begin{tabular}[c]{@{}c@{}}Improved\\ performance\end{tabular}                                                  \\ \hline \hline
\cite{GenISAC5} & GAN-CBD                                                                                                    & CSI                                                           & \begin{tabular}[c]{@{}c@{}}MSE of the estimated \\ channel and true channel\end{tabular}                            & \begin{tabular}[c]{@{}c@{}}Normalized MSE of  estimated channel\\  and true channel\end{tabular}                \\ \hline
\cite{GenISAC6} & GAN                                                                                                        & CSI                                                           & \begin{tabular}[c]{@{}c@{}}Wasserstein distance of generated \\ channel and true channel distributions\end{tabular} & \begin{tabular}[c]{@{}c@{}}Achievable rate, PDF of average\\  singular value of reflected channels\end{tabular} \\ \hline
\cite{GenISAC1} & \begin{tabular}[c]{@{}c@{}}Encoder-decoder based transformer \\  with auxiliary discriminator\end{tabular} & \begin{tabular}[c]{@{}c@{}}User \\ association\end{tabular}   & Handover  cost                                                                                                      & \begin{tabular}[c]{@{}c@{}}Spectral efficiency, Reliability and resilience,\\ QoPE of XR users\end{tabular}     \\ \hline
\end{tabular}
\end{table*}
Table~\ref{tab_ISAC} summarizes the application of GAI for optimizing ISAC-enabled network performance across different scenarios in \cite{GenISAC1, GenISAC5, GenISAC6}. Based on it, in \cite{GenISAC5}, GAI, through a GAN-CBD, offers an approach to improve signal processing by denoising data. Unlike traditional methods, GAN-CBD does not require prior knowledge of specific noise types. It tackles noisy channel data, a significant challenge for accurate CSI estimation. GAN-CBD estimates noise levels and performs non-blind denoising by employing a two-stage subnetwork architecture. 

In \cite{GenISAC6}, the limitations of traditional GANs for CSI estimation in IRS-aided communication systems is addressed. The proposed IRS-GAN framework leverages model-driven DL, incorporating prior knowledge about channel properties to achieve faster and more stable training. The GAI in this framework captures spatial correlations and user-specific channel characteristics. It employs a Wasserstein distance loss function \cite{WesGAN}, promoting training stability and convergence. Additionally, IRS-GAN demonstrates efficient scalability to multi-user scenarios. A single generative model can learn reflected channels for all users simultaneously, reducing network complexity compared to using separate GANs.  Finally, by enabling parallel training for its multiple discriminative models (one per user), the framework achieves further acceleration of the overall training process. Generative AI (GAI), as demonstrated in studies \cite{GenISAC5, GenISAC6}, can optimize networks by providing accurate CSI. This enables better decision-making regarding other network parameters, ultimately leading to enhanced network performance.

In \cite{GenISAC1}, addresses the issue of incomplete sensing information by employing a DL model to predict missing values in sensing data. This leads to comprehensive and continuous information about user behavior and environmental dynamics. By leveraging this richer data set, GAI optimizes network parameters like user association. It enables proactive handover decisions through an encoder-decoder transformer structure, minimizing disruptions for users.

\section{GAI-Based Optimization in SemCom-Enabled Networks}\label{SemComOpt}

\subsection{GAI in SemCom-Enabled Networks}

 In wireless communications, SemCom represents an emerging paradigm that operates on the innovative concept of semantic-meaning passing \cite{GenSemCom2}. At its core, SemCom involves extracting meanings from transmitted information at the transmitter. This process is facilitated by a matched  knowledge bases (KB) between the transmitter and the receiver, i.e., provisioning massive data samples to serve diverse AI learning and prediction tasks \cite{GenSemCom3}. As a result, AI plays a pivotal role in enabling SemCom.

\subsubsection{GFlowNet for intent-based SemCom}
 For the intent-based SemCom framework developed in \cite{GenSemCom4}, the smart system performs three key functions: measuring confidence in information, enhancing reasoning abilities in devices, and optimizing information transmission and reception. Confidence in the knowledge base (KB) is assessed using a probabilistic method based on fuzzy semantics. Here, the concept of Neuro-Symbolic AI (NeSy AI) plays a crucial role. NeSy AI leverages the strengths of both NNs and symbolic AI to create more powerful, flexible, and interpretable intelligent systems. NNs excel at pattern recognition and learning from data, while symbolic AI relies on well-defined knowledge representations and logic rules. The combination allows NeSy AI to provide deeper insights into its decision-making processes and potentially learn from smaller datasets compared to pure neural networks. This is particularly advantageous in scenarios with limited data availability or high data collection costs. Consequently, NeSy AI is well-suited for tackling problems that require both perception (understanding the environment through sensors) and reasoning (combining knowledge and logic).

NeSy AI in \cite{GenSemCom4} is employed to empower end nodes with reasoning capabilities, making them more intelligent. In pursuit of selecting the most suitable message for transmission, the objective function is formulated based on parameters for the GFlowNet. These parameters account for physical channel effects, imperfections in the KB at the destination node, and the causal structure learning of the data generator (Bayesian network). The objective is to minimize the causal influence on SemCom capturing the causal impact
of the sender’s message as observed through a channel. This objective also quantifies the discrepancies between the KB of the sender and that of the receiver. This problem is subjected to the constraint of semantic reliability which is given in terms of the squared error between the sender’s transmitted message and the receiver’s learned message. It is noteworthy that the optimization problem aims to identify optimal encoding and decoding methods, achieved through GFlowNet training.
 Acting as a NN-based brain, GFlowNet manages encoding and decoding processes efficiently. As mentioned in \cite{GenSemCom4}, the objective function of the problem is convex and by backpropagation to fine-tune the GFlowNet and encoder/decoder DNN weights, the unique optimal solution is obtained.  

\subsubsection{GDM-Based models for covert communication in SemCom systems}
Joint training of encoder-decoder pairs in SemCom systems can lead to significant computational overhead, which poses a considerable challenge for deploying network devices. However, the advanced learning capabilities of GAI offer a solution by allowing semantic decoders to reconstruct source messages using limited semantic information, such as prompts. While this approach eliminates the need for joint training with semantic encoders, the diverse abilities of GAI may introduce instability in the output. To address this, multi-modal prompts, including textual descriptions and structural information from images, are utilized in \cite{GenSemCom5}. These prompts are extracted at the semantic encoder stage. The semantic decoder then utilizes this information to generate the source image. The network consists of a transmitter, a receiver, a friendly jammer, and a warden in an open wireless environment. This work employs covert communication technique aiming to hide the transmission behaviour. The warden is presented with two possibilities: either the transmitter is inactive or actively sending a message. By evaluating factors like transmit power, jamming power, and signal degradation over distance, the warden attempts to make a decision. However, the success of covert communication hinges on misleading the warden. This is achieved by maximizing the detection error probability (DEP), which reflects the warden's likelihood of making incorrect decisions (false alarms or missed detections). A high DEP signifies the warden's confusion, making it difficult to determine if a secret message is being transmitted. Covert communication is successful when the DEP exceeds a threshold. Both the semantic encoder and decoder employ GDM-based models to extract prompts and generate images. Increasing the number of steps in the diffusion-based model can enhance robustness against noise but may lead to excessive energy consumption. To strike a balance between the number of diffusion steps required for extracting/generating images and energy consumption, joint GDM-based optimization is performed in \cite{GenSemCom5}. The details are given in what follows.

The transmitter's objective is to transmit images to the receiver while evading detection by a warden, achieving covert communication. This necessitates concealing the very existence of the transmission. To achieve this, both the transmitter and receiver leverage GDM-based models: the transmitter for prompt extraction and the receiver for image generation. The primary goal is to maximize the structural similarity between the original and reconstructed images. Achieving this objective is subjected to the constraint of ensuring that communications remain covert. Additionally, the total power consumption, encompassing the transmitter, jammer, and computational steps of the GDM model, must be below a predefined threshold.

To address this challenge, a two-stage GDM-based algorithm is employed. The first stage introduces a \emph{condition vector} that encapsulates various factors influencing the optimal resource allocation scheme. These factors include distances between the transmitter/jammer and the warden, path loss exponents for their respective communication links, and small-scale fading effects. This condition vector informs a scheme evaluation network that predicts the effectiveness of a given resource allocation scheme for a specific scenario. The second stage leverages a scheme \emph{generation network}, trained to generate resource allocation schemes that maximize the image reconstruction quality while maintaining covertness (high DEP). The collaboration between the stages approach allows the generation network to learn efficient resource allocation strategies for diverse conditions.

\subsubsection{GAN-based architecture for enhancing message interpretation in SemCom}
The work in \cite{GenSemCom7} proposes an architecture for implicit semantic-aware communication (iSAC).  iSAC allows the intended (destination) user to recognize and interpret hidden information within a message, such as hidden relations, concepts, and implicit reasoning processes, that are not directly conveyed by the source signal. The architecture operates in two phases: training and communication. The details are given in what follows.

Based on the system model developed in \cite{GenSemCom7}, the \emph{source user side} consists of an explicit semantics detector, explicit semantic encoder, and semantic comparator. Explicit semantics detector is an encoder that first identifies basic elements in the signal, such as objects or spoken words. However, these labels only capture a surface level of meaning. To truly grasp the full message, these labels should be linked to additional factors like hidden features and relationships between the identified elements. Explicit semantic encoder compresses the high-dimensional data into a semantic constellation space suitable for physical channels. The study in \cite{GenSemCom7} aims to achieve both compression and robustness in a single step. The function can handle joint or separate encoding of entities and relations, significantly reducing the data dimension for transmission. Semantic comparator focuses on how well the intended meaning is conveyed.  A key metric is semantic distance, which measures the difference between the sender's intended meaning and the receiver's interpretation.  This study proposes a method to evaluate this distance by comparing the semantic reasoning paths inferred from the transmitted information. The paths represent the broader understanding derived from the message's explicit elements. For \emph{communication channel}, a limited-capacity channel with noise and fading is considered. The received signal is a combination of the encoded message, the channel's clarity, and background noise. Since the sender does not know how clear the channel is, the receiver needs to interpret the meaning from the noisy version of the transmitted information. The \emph{receiver} utilizes a semantic interpreter to understand the intended meaning behind the transmitted information. This interpreter builds and analyzes possible reasoning paths. It can identify likely hidden relations and entities that the sender might have intended. The interpreter continuously learns and improves its reasoning abilities based on the data it receives.

Given the system model above, the primary objective in \cite{GenSemCom7}  is to develop a system that empowers the receiver to automatically infer the intended meaning behind the transmitted information. This is particularly challenging because communication messages can often contain hidden layers of meaning and relationships between elements. To address this complexity, the problem problem to minimize the difference (semantic distance) between the sender's intended meaning paths and the receiver's estimated meaning paths. However, due to the complexity of real-world communication, a straightforward solution is impractical. Therefore, a \emph{generative imitation learning-based}  framework is derived, allowing the receiver's semantic interpreter to learn from successful examples and progressively improve its ability to infer the intended meaning from the received information. In the developed iSAC architecture for addressing the aforementioned problem, the details for semantic encoding, semantic distance, semantic comparator at \emph{user side}, and semantic interpreter at \emph{destination side} are given in what follows.

 The \emph{semantic encoding} scheme has two crucial properties: efficiency and robustness. The encoding process utilizes a projection-based function to compress the semantic information into a lower-dimensional semantic constellation space, enabling efficient transmission through physical channels. To enhance robustness against channel corruption, the system leverages the sender's past communication patterns. Frequently used combinations of entities and relations are preferred and encoded in a way that makes them distinct from less frequent combinations. This allows the receiver to make guesses about the intended meaning even if some elements are corrupted during transmission. The projection function can be pre-trained by the sender and shared with the receiver beforehand, eliminating the need for the receiver to possess knowledge of the sender's specific communication history. For measuring the \emph{semantic distance} between the intended and inferred meanings, a statistic-based distance measure is proposed that focuses on the reasoning process. The sender shares example paths, and the receiver utilizes them to train its own reasoning mechanism to mimic their behavior. The semantic distance is then measured by comparing the probabilities of taking different reasoning paths between the expert and the receiver's model. And, the problem of minimizing the semantic distance between the sender's intended meaning paths and the receiver's estimated meaning paths is reformulated as minimizing this Jensen-Shannon (JS) divergence-based semantic distance. This problem would bridge the gap between the sender's intended meaning and the receiver's interpretation.

 Since the destination user cannot directly observe the expert semantic paths, providing effective feedback for training its reasoning mechanism presents a challenge. Sharing the true probability distributions of reasoning paths, is impractical as these depend on the unknown expert reasoning process. To address this, \emph{semantic comparator} utilizing a discriminator network is developed. This network is trained to distinguish between expert semantic paths from the source user and the paths inferred by the destination user's reasoning mechanism. As shown in \cite{GenSemCom7}, the network's ability to perform this discrimination is mathematically equivalent to the true semantic distance between the reasoning mechanisms. This eliminates the need for revealing any confidential information about the expert paths or probabilities. The source user can simply use the output of the discriminator network to measure the semantic distance and guide the destination user's model correction and training. This feedback loop allows the destination user to progressively improve its reasoning mechanism, ultimately leading to a better understanding of the implicit meaning conveyed in the messages.

The main component at the destination side is the \emph{semantic interpreter}, aiming to minimize the semantic distance between the expert semantic paths and the interpreted paths generated by the policy network of the destination user. In \cite{GenSemCom7}, for the destination user, the implicit semantic reasoning process from the received explicit semantics is formulated as a MDP problem. The MDP developed in \cite{GenSemCom7} guides the destination user in inferring implicit meaning. The state space reflects progress in understanding by considering various reasoning paths of different lengths.  Actions involve choosing the next relation to extend a path. The reward function incentivizes the generation of paths semantically close to the source's intended meaning, even without explicit expert paths. The system learns it through indirect feedback and optimizes for efficient transmission of generated paths.  Acting as the system's brain, the policy network determines the most promising relation to explore the next understanding based on the current one, ultimately guiding the system towards paths that align with the source's meaning.

During the training phase, the semantic interpreter at the destination side aims to generate reasoning paths that extend from the recovered explicit semantics, acting as an AI to understand the deeper meaning behind the message. Specifically, through a series of reasoning steps (episodes), the interpreter builds the reasoning path. In each episode, the interpreter considers the current state (based on the recovered explicit semantics) and chooses an action. This action involves selecting a set of relations (concepts linking entities) to extend the path further. After selecting relations, the interpreter updates the current reasoning path by adding the chosen relations and the newly linked entities. This process of selecting actions and updating the path repeats until the path reaches its maximum allowed length. The value of allowed length is determined based on the complexity of the message (depth of meaning) typically observed in past communication between the source and destination users. 

\subsubsection{GDM-based power control for Semantic UAV Communications}
GDM-based power control is proposed in \cite{GenSemCom6} to determine the portion of the transmit power level for a UAV that is assigned to an object to be transmitted to the user. The semantic features of each object are derived by an object detector. To maximize the total semantic transmission quality score for the objects, given in terms of their importance, the combination of the DRL and DDPM algorithms is employed. The output, which is denoised data, determines the transmission power weight for each object.

\subsubsection{GDM-based models for full-duplex D2D SemCom}
In \cite{AIGC2}, a full-duplex device-to-device (D2D) semantic communication scheme is proposed to support information sharing among multiple mixed-reality (MR) users. This scheme enables efficient synchronization of free-space and semantic information. To encourage MR users for semantic information sharing, the authors design an optimal contract using a diffusion model.
Within the proposed framework, semantic information is extracted from a user's view image. First, the image is fed into a semantic encoder that extracts interest points and descriptions. This encoder leverages the self-supervised SuperPoint architecture \cite{SupPoint}, which consists of a shared CNN for dimensionality reduction and separate encoders for interest points and descriptors. Extracted interest points and descriptions from other users are then received through D2D wireless communication for information matching.
The framework utilizes SuperGlue \cite{SupGlue} as a semantic matching network architecture. This architecture employs an attentional graph neural network to enhance the uniqueness of interest points and descriptors, followed by an optimal matching layer that generates a partial assignment matrix. Additionally, it is trained in a supervised manner using a negative likelihood loss function for the assignment matrix.
Finally, a payment plan is developed that incentivizes both the semantic information producer (SIP) and the semantic information receiver (SIR) to cooperate. The utility for the SIP considers the quality of the shared information and its transmit power level. The SIR's utility is based on the quality of the information received from both itself and the SIP. The plan aims to maximize the SIR's utility while satisfying a constraint that ensures the SIP's utility is met. A diffusion-based model (i.e. DDPM) is used to solve this problem, ultimately deriving an AI-generated contract.

\subsubsection{GDM-based framework for integrated SemCom and AIGC in metaverse applications} 
The study \cite{AIGC6} explores the potential of a unified framework for integrated SemCom and AIGC (ISGC) within metaverse applications. This framework aims to enhance both communication and content generation in the metaverse, including semantic, inference (AIGC), and rendering modules. ISGC offers several key functionalities including generating visually appealing and  highly relevant content to the user's context and needs. Its adaptability to the user's context and needs can address two major challenges in the Metaverse such as inefficient resource utilization and low-quality content.   

Based on the findings in \cite{AIGC6}, the problem of bandwidth allocation in the metaverse is expressed as a MDP to maximize utility for metaverse service providers (MSPs). The state space encompasses various parameters relevant to SemCom performance, including semantic entropy, average transmitted symbols from the semantic module, and channel gains. Additionally, it incorporates transmit power and channel gain from the semantic module, and those  the inference module to the rendering module. Furthermore, the state space accounts for computing resources and additive Gaussian noises present at both the AIGC and rendering modules. The action space encompasses the available bandwidth allocation from the semantic, AIGC, and rendering modules, respectively.  Finally, a reward function is defined based on the utility for the MSP. To solve this MDP and achieve optimal bandwidth allocation, a combination of a diffusion model and DRL is employed. The details are given the following.

At each step, the method commences by observing the current state. This information serves as the initial input. Subsequently, Gaussian noise is injected into the system to initialize candidate actions. The reverse diffusion process is then employed to refine these actions, effectively removing the noise and generating suitable control signals. To encourage exploration within the environment, exploration noise is further added to the resulting actions. Following action execution, the corresponding reward, determined by a pre-defined utility function, is obtained. Additionally, the complete environment record, encompassing both state and reward information, is stored within a replay buffer for future training iterations. To further enhance the model's performance, a mini-batch of records is randomly sampled from the replay buffer. Leveraging these samples, the critic networks are updated by computing the loss function and applying the policy gradient. Finally, the target networks are updated to synchronize their parameters with the critic networks. 

\subsection{How GAI Models are Used for SemCom-Enabled Network Optimization?}
\begin{table*}[]
\caption{Optimizing SemCom-enabled networks with GAI}
\label{tab_SemCom}
\begin{tabular}{|c||c||c||c||c|}
\hline
Ref.       & GAI model                                                                & Optimizing variable                                                                                                                         & Objective/loss function                                                                                                        & Improved performance                                                                                                                                                                                         \\ \hline \hline
\cite{GenSemCom4} & GFlowNet                                                                 & \begin{tabular}[c]{@{}c@{}}DNNs' parameters for encoder,\\ decoder, and GFlowNet\end{tabular}                                               & Causal influence on SemCom                                                                                                     & \begin{tabular}[c]{@{}c@{}}Amount of transmitted bits,\\ Semantic reliability\end{tabular}                                                                                                                    \\ \hline
\cite{GenSemCom5} & GDM                                                                      & \begin{tabular}[c]{@{}c@{}}Transmitter and warden power,\\ number of steps for diffusion \\ model\end{tabular}                             & \begin{tabular}[c]{@{}c@{}}Structural similarity of the  \\ source and receiver images \\ (SSM)\end{tabular}                   & Detection error probability, SSM                                                                                                                                                                             \\ \hline
\cite{GenSemCom7} & \begin{tabular}[c]{@{}c@{}}Generative imitation \\ learning\end{tabular} & \begin{tabular}[c]{@{}c@{}}Parameters of the semantic\\ interpreter at the destination user\end{tabular}                                   & \begin{tabular}[c]{@{}c@{}}JS divergence-based semantic\\  distance between the intended and \\ inferred meanings\end{tabular} & \begin{tabular}[c]{@{}c@{}}Implicit meaning interpretation \\ at destination user\end{tabular}                                                                                                               \\ \hline
\cite{GenSemCom6} & GDM                                                                      & \begin{tabular}[c]{@{}c@{}}Portions of the UAV transmit \\ power level\end{tabular}                                                        & \begin{tabular}[c]{@{}c@{}}Total semantic transmission \\ quality score for the objects\end{tabular}                           & Transmission quality, Bit error rate                                                                                                                                                                          \\ \hline
\cite{AIGC2}      & GDM                                                                      & \begin{tabular}[c]{@{}c@{}}Transmit power level for SIP, \\ Contract parameters\end{tabular} & SIR utility                                                                                                                    & \begin{tabular}[c]{@{}c@{}}Bit error probability for a user,\\ Achievable rate for a user, \\ Contract values \end{tabular} \\ \hline
\cite{AIGC6}      & GDM                                                                      & Bandwidth                                                                                                                                  & Utility for the MSPs                                                                                                           & \begin{tabular}[c]{@{}c@{}}Bandwidth utilization for \\ enhancing user experience\end{tabular}                                                                                                               \\ \hline
\end{tabular}
\end{table*}
The application of GAI for optimizing SemCom network performance across different scenarios in \cite{GenSemCom4, GenSemCom5, GenSemCom6, GenSemCom7, AIGC2, AIGC6} is summarized in Table~\ref{tab_SemCom}. 
By leveraging NeSy AI \cite{GenSemCom4}, GAI empowers devices (end nodes) with reasoning capabilities, enabling them to grasp the context and implications of information for more efficient and targeted communication. Furthermore, NeSy AI facilitates the selection of optimal messages for transmission by considering channel effects, potential KB discrepancies between sender and receiver, and the causal structure of the transmitted data. This allows for the minimization of misunderstandings and ensures the intended causal impact on the receiver. The framework also tackles the trade-off between semantic reliability, minimizing message distortion, and transmission efficiency, using a specifically designed objective function to train the GFlowNet. Notably, NeSy AI's ability to learn effectively from limited data makes it advantageous in scenarios where data collection is restricted.

The focus of study in \cite{GenSemCom5} is on covert communication, where the goal is to transmit messages undetected by a monitoring entity. To ensure covertness, the system prioritizes maximizing the DEP, which signifies the warden's confusion regarding the presence of a hidden message. A GDM-based optimization is implemented to strike a balance between achieving high DEP (covertness) and minimizing energy consumption by the transmitter, jammer, and the GDM models themselves. GAI model in this study addresses the  computational efficiency, stability, covertness, and energy usage.

In \cite{GenSemCom7}, to optimize networking for implicit semantic-aware communication, a GAI model empowers the receiver to grasp the hidden meaning behind messages, encompassing relationships and concepts that go beyond what is explicitly conveyed. GAI facilitates receiver learning of implicit meaning via generative imitation learning. The receiver's semantic interpreter progressively improves its ability to infer hidden meaning by learning from successful examples (expert paths). A discriminator network, also powered by GAI, distinguishes between expert and receiver-generated paths, revealing the semantic distance without sharing confidential information. This feedback guides the receiver's model correction and training. GAI enables learning from indirect feedback through a reward function that incentivizes the generation of semantically close paths. By incorporating these GAI-powered techniques, the system fosters a more robust understanding of the information being communicated.

In the context of MR user communication explored in \cite{AIGC2}, GAI optimizes networking for efficient semantic information sharing through AI-generated contract. A critical challenge in information sharing is ensuring cooperation between users. This study addresses this by employing DDPM to design an optimal contract. This GAI-powered contract incentivizes both SIP and SIR to cooperate by considering factors like transmission power for the SIP, and contract parameters. The DDPM iteratively refines the contract terms to find a solution that maximizes the SIR's utility while ensuring the SIP's needs are met. 

In the context of metaverse applications explored in \cite{AIGC6}, GAI optimizes network parameters for ISGC-enabled network. To achieve optimal bandwidth allocation within this complex MDP, the study utilizes a combination of a diffusion model and DRL. The diffusion model acts as a refinement tool. It injects noise into candidate actions and then employs a reverse diffusion process to remove the noise, effectively guiding the system towards suitable control signals for bandwidth allocation.

\section{GAI-Based Optimization in xG Communications Security}\label{xGSecurityOpt}

\subsection{GAI in xG Communications Security}
In xG networks, GAI would be a promising tool to ensure robust security communication. This section delves into recent research exploring how GAI can be leveraged to address security challenges in xG communication. 

\subsubsection{GANs for intelligent intrusion detection in IoT-enabled networks}
Securing xG wireless networks that employ Internet-of-Things (IoT) devices for edge computing presents distinct challenges. Limited computing and storage capacities at edge nodes, reliance on open networks like satellites and WiFi for data transmission, and the inability of existing intrusion detection systems (IDSs) to identify novel attacks (e.g., new Distributed Denial-of-Service attacks) are key concerns. To address these issues, a solution for intelligent intrusion detection in IoT-based edge computing is proposed in \cite{GenSec2}. This scheme leverages fuzzy rough sets for efficient and rapid feature extraction from data, coupled with a deep convolutional GAN to augment training samples.

Feature selection in intrusion detection systems is a multi-step process. The first step involves \emph{information processing}, where raw network data from edge nodes undergoes preprocessing (e.g., using CICFlowMeter) to generate standardized features that aid in attack data identification without disrupting normal traffic. The second step focuses on \emph{choosing the proper features}. Here, a decision system denoted as $S = (U, C \bigcup D_x, V)$ is employed. Within this system, $U$ represents the set of collected data items, $C$ represents the set of characteristics describing these items, $D_x$ represents the label assigned to each item in $U$, and $V$ represents the set of all possible values for both the features in $C$ and the decision labels in $D_x$. This phase leverages a two-step approach. First, the membership degree of all elements within the data set $U$ is calculated (as described in \cite{GenSec2}). Then, to select relevant features from the set $C$, two metrics are evaluated for each feature by using the membership degrees: decision loss and the difference between the individual feature and the entire feature set $C$. These metrics leverage the membership degrees of the elements in $U$. This difference is quantified using the fuzzy knowledge distance (FKD) metric, as detailed in \cite{FKD}. If both the decision loss and the FKD value for a particular feature exceed predefined thresholds, that feature is selected for further analysis.

While feature selection can improve intrusion detection efficiency, it can also lead to data loss and potentially impact accuracy. To address this, \cite{GenSec2} incorporates feature correlation modeling. This approach leverages the FKD metric to analyze the relationships between features within the selected data set. By constructing a feature correlation mechanism based on FKD, features can be prioritized for efficient extraction, ultimately enhancing the accuracy of network intrusion detection.

Benefiting from the power of convolutional neural networks (CNNs) combining CNNs with a GAN, intrusion detection is then implemented. By generating synthetic attack data, this scheme can potentially achieve similar training results with a smaller actual dataset, which would be more suitable for edge nodes. This enables the system to handle diverse network scenarios.


The study in \cite{GenSec4} introduces an AI-based network IDS that leverages the boundary equilibrium GAN (BEGAN), to address critical data imbalance issues inherent in network traffic. This system enhances detection capabilities by generating synthetic attack data, thus enriching the training sets and enabling the detection models to learn a wider array of attack behaviors, including those that are rare and complex. The four-stage architecture of the system- preprocessing, generative model training, autoencoder training, and predictive model training- ensures data processing and feature extraction, boosting the accuracy and reliability of threat detection. The details are explained in what follows.

The first stage, \emph{preprocessing}, focuses on refining the raw data set. This involves three subprocesses: outlier analysis, one-hot encoding, and feature scaling. Outlier analysis utilizes the median absolute deviation (MAD) measure to identify and remove data points that deviate significantly from the expected normal distribution of the attributes. The standard deviation is then calculated based on the MAD and used as a threshold for outlier filtering. Next, the remaining attributes are transformed into one-hot vectors. This encoding method represents each categorical attribute value (e.g., protocol values like TCP, UDP, and ICMP) as a separate binary feature within a vector. The size of the vector corresponds to the total number of possible attribute values. Finally, the system scales the numeric attributes using the min-max normalization method.

Following preprocessing, this refined data is used to \emph{train a generative model} called BEGAN. This model builds upon the concept of autoencoders, which are neural networks that learn to compress and reconstruct data. BEGAN utilizes a reconstruction error-based objective function, meaning it aims to minimize the difference between the original data and the data it reconstructs.
The BEGAN model's discriminator architecture is based on a symmetrical autoencoder with five layers. The generator, in turn, mirrors the decoder portion of this autoencoder architecture. In BEGAN, multiple generative models are built, typically one for each class of data. After training, each generative model specializes in producing synthetic data that corresponds only to its assigned class. BEGAN offers a unique advantage during training. Unlike some generative models, BEGAN can reach a state of equilibrium, where the generator and discriminator are evenly matched. This characteristic simplifies the training process by making it easier to determine when to stop training. 

After the preprocessing and generative model training phases, the system leverages the expanded training set for the final two stages: autoencoder training and detection model training. The autoencoder, a type of NN, is trained to learn compressed representations of the network traffic data. This compressed data can be beneficial for dimensionality reduction, which can improve the efficiency of training the subsequent detection model.

For the detection model, the system offers flexibility in choosing from various architectures, including basic DNNs, CNNs, and LSTM networks. Each architecture is suited to extract different features from the network traffic data. Basic DNNs provide a good general-purpose option, CNNs excel at extracting spatial features, and LSTMs are particularly adept at analyzing sequential data with temporal dependencies. The choice of architecture depends on the specific characteristics of the network traffic data and the desired outcome of the detection model.


In IoT networks assisted by UAVs, as the number of UAVs increases, the risk of cyberattacks also rises. Additionally, some UAVs suffer from limited local data samples and imbalanced data distribution, hindering the efficiency of local model updates and the robustness of IDS models. To address these challenges, \cite{GenSec5} proposes an IDS that uses conditional GANs to effectively learn the data distribution and guide balanced data generation. To address the challenges of vanishing gradients, retaining long-term contextual information, and capturing subtle variations between normal and attack data, the IDS employs long short-term memory (LSTM) networks in both the generator and discriminator of the conditional GAN. The LSTM network in the generator leverages its ability to capture fine-grained features of sequential data to generate new samples based on the corresponding conditional labels. These generated samples are then fed into the discriminator alongside real, labeled data, where the LSTM within the discriminator attempts to distinguish between them. This adversarial training process, where the generator aims to produce increasingly realistic data and the discriminator strives to identify forgeries accurately, iterates alternately. Ultimately, the trained LSTM network within the discriminator becomes a powerful classifier, enabling the IDS to detect and accurately classify attack data. By leveraging the feature extraction ability of LSTMs, the GAN-generated data acts as augmented data, enhancing intrusion detection and classification performance. 

While the conditional GANs-based method explained above can effectively learn data representations from the limited datasets collected by individual drones, these representations may be biased due to variations in data type, volume, and collection methods across UAVs. Furthermore, encountering a new attack type necessitates retraining each drone with the new data, a time-consuming process. In \cite{GenSec5}, instead of sharing data directly between drones, distributed FL is employed, providing a more efficient and secure alternative by enabling UAVs to collaboratively train a model by sharing only the trained model parameters, not the raw data itself. Specifically, a novel approach for collaborative intrusion detection is proposed in \cite{GenSec5}. This framework leverages a blockchain-powered distributed FL architecture to enhance performance. It operates in two stages: local model training at the UAV level and global model aggregation at the MEC level. The details are given in what follows.

During the \emph{local training stage}, each UAV, acting as a federated user, trains its own IDS model based on conditional GANs using its local data. To preserve data privacy, Gaussian noise is injected into both the data features and the local model parameters before transmission. In the \emph{global model aggregation stage}, a mobile edge computing (MEC) node acts as the blockchain maintainer. It collects the trained IDS model parameters from all UAVs and stores them as blockchain transactions. The MEC node then aims to minimize a global loss function, which is given in terms of the combined loss function from all UAVs and the sum of their model parameters. This distributed approach utilizes multiple MEC nodes for collaborative training, fostering robustness through a secure blockchain network. The blockchain stores the evolving global model and facilitates consensus among nodes on its validity, eliminating single points of failure and ensuring data integrity. Additionally, MEC nodes with better performing models contribute more significantly to the global model, incentivizing participation and promoting overall accuracy.

To achieve a continuously improving global intrusion detection model, a trust-based update process is employed. MEC nodes calculate trust values for each UAV. When a trust value exceeds a predefined threshold, the MEC node receives the local model parameters from the UAV. These local parameters are then aggregated across all UAVs. The resulting global parameters are subsequently broadcast back to each UAV, where they are incorporated into the local model training process. This iterative process of local training, parameter aggregation, and global model distribution continues until the overall model converges, enabling accurate detection and classification of attack data.

\subsubsection{GAN-based models for trust management in industrial wireless sensor networks}
The dynamic nature of industrial wireless sensor networks (IWSNs) demands robust security. Hence, a trust management scheme is crucial to distinguish legitimate data from malicious attacks and tolerate network issues for reliable communication. As demonstrated in \cite{GenSec10}, such a scheme can be developed with functionalities like trust evidence collection, trust classification, and trust redemption. 

For \emph{trust evidence collection}, the IWSNs should have the ability to distinguish between trustworthy and malicious nodes. In doing so, \cite{GenSec10} focuses on collecting evidence regarding data transmissions, such as packet drops, delays, and tampering attempts. However, the industrial environment can introduce uncertainties – node faults or interference can mimic attacks. To account for this ambiguity, the scheme leverages Interval Type-2 Fuzzy Logic, exceling at handling both fuzziness and randomness inherent in the collected evidence. The scheme calculates trust attributes like packet loss rate and transfer delay rate, and utilizes fuzzy sets to categorize these attributes into low, medium, or high trust levels. Furthermore, predefined fuzzy rules map these trust attributes to a final trust value for each node. This approach allows the system to make robust trust evaluations even in the presence of uncertainties within the IWSN. 

Traditional trust classification methods struggle in industrial applications due to limited experience and the dynamic environment. Inspired by GANs' ability to learn sample distributions, the authors in \cite{GenSec10} propose a encoder-decoder structure for \emph{trust classification}. Trust values from the initial deployment (assumed to have few malicious nodes) are used to create training data (real samples) in the form of trust vectors (sequences of historical trust values).  A conditional GAN acts as the decoder, while a standard GAN serves as the encoder. The encoder learns a compressed representation (latent data) of a trust vector. The decoder reconstructs a trust vector based on the latent data and conditional information (changes in trust values). Both the GAN and CGAN are optimized through adversarial training. Batch normalization is employed to improve training stability and convergence\footnote{Batch normalization tackles exploding or vanishing gradients, a problem in DNNs. It normalizes layer outputs during training, leading to faster convergence and more stable training.}. After training, the current trust vector of a node is fed into the encoder-decoder structure. The reconstruction loss is then used to determine whether the node is trustworthy or not.

For \emph{trust redemption}, it is critical to mitigate the effects of false positive malicious node detection. In \cite{GenSec10}, a GAN-based trust redemption model is proposed. This model allows potentially good nodes, mistakenly identified as malicious, to regain trust within the network. During steady operation, the system gathers sequences of trust evidence from labeled malicious nodes. These sequences are then processed to create attack probability vectors, forming the training data for the GAN. The GAN's generator attempts to learn the features of real attack probabilities and predict missing information. The GAN's discriminator, on the other hand, aims to distinguish between genuine and generated data. Through an iterative training process, the model refines its ability to predict attack probabilities. Subsequently, this trained model can assess future attack likelihood for suspect nodes. If the model predicts a low probability of malicious behavior, the node can potentially be reintegrated into the network, improving overall network efficiency.

In \cite{GenSec10}, GANs are leveraged to address critical security challenges in IWSNs because traditional classification methods often struggle due to limited data and the dynamic nature of these networks. In this study, the GAN-based encoder-decoder structure analyzes historical trust data and distinguishes trustworthy nodes from potential attackers, effectively handling uncertainties inherent in the network. Moreover, GANs play a key role in the trust redemption model, which mitigates the effects of false positives. By analyzing evidence from labeled malicious nodes, the model can predict the likelihood of future attacks. This allows the network to potentially reintegrate good nodes that were mistakenly flagged, leading to a more robust and efficient IWSN security system.  

\subsection{How GAI Models are Used for Network Security Optimization?}
\begin{table*}[]
\caption{Security in xG wireless networks with GAI}
\label{tab_Security}
\begin{tabular}{|c||c||c||c||c||c|}
\hline
Ref.     & GAI model                                                           & Objective           & \begin{tabular}[c]{@{}c@{}}Methodology\\ for enhancement\end{tabular}                         & Challenge                                                                                                                 & \begin{tabular}[c]{@{}c@{}}Network \\ Improvement\end{tabular}                                                                      \\ \hline \hline
\cite{GenSec2}  & \begin{tabular}[c]{@{}c@{}}Deep convolutional\\ GANs\end{tabular}   & Intrusion detection & \begin{tabular}[c]{@{}c@{}}FKD for feature\\ selection\end{tabular}                           & \begin{tabular}[c]{@{}c@{}}Data scarcity and  model training \\ efficiency at the network's edge\end{tabular}             & Accuracy                                                                                                                            \\ \hline
\cite{GenSec4}  & Wasserstein GANs                                                    & Intrusion detection & \begin{tabular}[c]{@{}c@{}}Autoencoder for\\ anomaly detection\end{tabular}                   & \begin{tabular}[c]{@{}c@{}}Data imbalance in network traffic \\ and anomaly detection\end{tabular}                        & Accuracy                                                                                                                            \\ \hline
\cite{GenSec5}  & \begin{tabular}[c]{@{}c@{}}Conditional\\ GANs-LSTM\end{tabular}     & Intrusion detection & \begin{tabular}[c]{@{}c@{}}FL and blockchain\\ for data integrity\end{tabular}                & \begin{tabular}[c]{@{}c@{}}Collaborative learning in \\ dynamic UAV environments with\\  emphasis on privacy\end{tabular} & Accuracy                                                                                                                            \\ \hline
\cite{GenSec10} & \begin{tabular}[c]{@{}c@{}}GANs-encoder \\ and decoder\end{tabular} & Trust management    & \begin{tabular}[c]{@{}c@{}}Type-2 fuzzy \\ logic systems for \\ trust evaluation\end{tabular} & \begin{tabular}[c]{@{}c@{}}Trust management in dynamic\\ and potentially hostile industrial \\ environments\end{tabular}  & \begin{tabular}[c]{@{}c@{}}Network lifetime,\\ Network throughput,\\ Packet dropping attack,\\ Packet delaying attack.\end{tabular} \\ \hline
\end{tabular}
\end{table*}
As discussed previously through studies \cite{GenSec2} and summarized in Table~\ref{tab_Security},  GAI can optimize security in xG communication.

The proposed GAI-based security schemes in  \cite{GenSec2, GenSec4, GenSec5} aim for a proactive security posture which is an objective for security in wireless networks, especially given the dynamic and often vulnerable nature of wireless communications. It refers to a strategic approach in cybersecurity where measures are implemented to prevent security incidents before they occur, rather than merely responding to them after they happen. This proactive stance involves anticipating potential threats and vulnerabilities and taking preemptive actions to mitigate or eliminate risks. In these studies, this objective is obtained by generating synthetic attack data. GAI expands the range of scenarios that the IDS can experience and learn from, ensuring that emerging threats can be detected early and accurately. 

In \cite{GenSec2}, FKD plays a critical role in streamlining the data that the IDS needs to process. By focusing on the most relevant features, it reduces the computational complexity and speeds up the detection process. To efficiently learn representations (encoding) of the data while reducing dimensionality, autoencoder is used in \cite{GenSec4}. This compressed feature representation helps in focusing on the most significant aspects of the data, enhancing the detection of anomalies by reducing the influence of noise and irrelevant information. Anomalies are detected based on the reconstruction error, which is the difference between the original input and its reconstruction from the compressed representation. In UAV networks in \cite{GenSec5}, where data may have temporal dependencies, LSTM is used effectively to capture these dynamics. It contributes to recognizing complex patterns and sequences in data, which are crucial for detecting sophisticated cyberthreats that may unfold over time. Additionally, FL allows multiple UAVs to collaboratively learn a shared intrusion detection model while keeping all the training data local to each UAV. This approach not only preserves the privacy of the data but also reduces the bandwidth needed for transferring large datasets. Blockchain builds trust in the system's integrity and the validity of shared data and model updates. Type-2 fuzzy logic in \cite{GenSec10} is used to evaluate the trustworthiness of each sensor node in the network dynamically. By continuously assessing the trust levels of sensor nodes, the system can make informed decisions about data credibility and node reliability.

\section{Network-Assisted Execution of GAI Models}\label{NetForGAI}
\begin{table*}[]
\caption{Analysis of networking enhancements for integration of GAI-based applications}
\label{NetForGAI}
\begin{tabular}{|c||c||c||c||c|}
\hline
Ref.   & Focus                                                                                                                                             & Networking role                                                                                                                                              & Contributions                                                                                                                                                               & Improvement                                                                               \\ \hline \hline
\cite{AIGC4}  & \begin{tabular}[c]{@{}c@{}}Managing AIGC lifecycle with \\ blockchain integration in edge \\ networks.\end{tabular}                               & \begin{tabular}[c]{@{}c@{}}Facilitating decentralized management \\ and secure transactions of AI content \\ between edge devices and networks.\end{tabular} & \begin{tabular}[c]{@{}c@{}}Blockchain for transaction \\ integrity, copyright protection, \\ and trust in AI content distribution.\end{tabular}                             & \begin{tabular}[c]{@{}c@{}}Blockchain\\ latency\end{tabular}                              \\ \hline
\cite{AIGC11} & \begin{tabular}[c]{@{}c@{}}Efficient AI content generation\\  using collaborative and distributed \\ computing in wireless networks.\end{tabular} & \begin{tabular}[c]{@{}c@{}}Enables collaborative processing \\ of AI tasks across distributed devices,\end{tabular}                                          & \begin{tabular}[c]{@{}c@{}}Distributed diffusion processes \\ to share the computational load \\ across networked devices.\end{tabular}                                     & \begin{tabular}[c]{@{}c@{}}Network resources \\ and \\ computational efforts\end{tabular} \\ \hline
\cite{AIGC7}  & \begin{tabular}[c]{@{}c@{}}Privacy-preserving training of AI \\ models for content generation using \\ FL in wireless networks.\end{tabular}      & \begin{tabular}[c]{@{}c@{}}Supporting the distribution of AI model \\ training across various clients without \\ central data aggregation\end{tabular}       & \begin{tabular}[c]{@{}c@{}}FL for decentralized AI \\ training across networked \\ devices, maintaining data privacy.\end{tabular}                                          & \begin{tabular}[c]{@{}c@{}}Privacy and BW \\ usage\end{tabular}                           \\ \hline
\cite{AIGC9}  & \begin{tabular}[c]{@{}c@{}}Secure and efficient AIGC in the \\ Metaverse using blockchain and \\ SemCom technologies.\end{tabular}                & \begin{tabular}[c]{@{}c@{}}Ffficiently handling large data transfers\\  required for AI content generation\end{tabular}                                      & \begin{tabular}[c]{@{}c@{}}Secure framework for transmitting \\ semantic data efficiently across \\ entities, using blockchain \\ and ZKPs for added security.\end{tabular} & Security                                                                                  \\ \hline
\end{tabular}
\end{table*}

Previously, the potential of the GAI models for optimizing the xG wireless networks was explored. In this section, as briefly discussed in Table~\ref{NetForGAI}, we expand the focus and delve into the role of efficient networking in facilitating the functionalities of GAI. 

\subsection{Blockchain-Enabled Lifecycle Management for Edge AIGC Products}
The recently proposed edge AIGC paradigm, where AIGC services are distributed to edge devices, effectively addresses latency issues but introduces significant challenges in managing the lifecycle of AIGC products. These challenges include tampering, plagiarism, and ensuring overall trustworthiness. To tackle these issues, a blockchain-powered framework is proposed in \cite{AIGC4} that comprehensively manages the lifecycle of AIGC products across three stages. In the first stage, generation, edge service providers (ESPs) utilize carefully crafted prompts to generate content using advanced AI models. Following creation, the AIGC products transition into the distribution stage, where they are disseminated across various platforms, ranging from popular social media to specialized AIGC platforms dedicated to sharing and consuming AI-generated content. The final stage, trading, involves the economic aspects of AIGC, typically modeled as transfers of ownership or rights between parties. The blockchain framework central to this approach involves multiple stakeholders, including producers, ESPs, consumers, and potential attackers. It provides a traceable and immutable ledger and supports essential on-chain mechanisms such as Proof of AIGC, incentive mechanisms, and ESP selection.

The proof of AIGC involves two critical phases: proof generation and challenge phase. During the former phase, each ESP registers its AIGC products on the blockchain, establishing ownership rights for the producers. The latter phase allows original content creators to protect their copyrights by identifying and challenging unauthorized copies on the blockchain, potentially leading to their deregistration.   

The Incentive Mechanism is designed to motivate all stakeholders to actively participate in lifecycle management by incorporating mechanisms like coinbase transactions, which reward block creators, and ensuring atomic execution of trades involving funds and ownership.

During ESP Selection, producers assess ESPs based on a reputation system that accounts for past interactions (both positive and negative) and includes a manually set uncertainty factor reflecting the quality of communication. This reputation system guides producers in choosing the most reliable ESP for their needs.

\subsection{Collaborative Distributed Diffusion for Energy-Efficient AIGC Execution} 
The focus of the study detailed in \cite{AIGC11} revolves around enhancing network functionalities to mitigate the limitations encountered during the deployment of AIGC. The study introduces a collaborative distributed diffusion-based AIGC framework designed to optimize computing energy usage and improve user experience by promoting device collaboration. This collaboration addresses the challenges faced by resource-constrained devices in executing AIGC services.

As discussed in Section~\ref{ModelGAI}, diffusion models such as DDPM progressively generate new data by iteratively removing noise from a latent representation. The study proposes three network architectures to support this process: edge-to-multiple devices, device-to-device, and multi-device clustering, which can function with or without an edge server's involvement.

In the edge-to-multiple devices architecture, an edge server serves as a central hub, processing shared denoising steps for groups of user devices tasked with semantically similar AIGC projects. Subsequently, each user completes the specific denoising steps independently, yielding benefits such as reduced latency, optimized resource allocation, and effective load balancing. The D2D architecture facilitates direct collaboration between two devices on AIGC tasks. Here, agreed-upon shared processing steps are performed by one device, after which intermediate results are exchanged, allowing each device to finalize the process independently. This approach promotes energy efficiency and enhances privacy. Furthermore, the clustering architecture involves groups of user devices collaborating to manage AIGC tasks. These clusters may be formed with the assistance of an edge server or through self-organization based on the capabilities of the devices involved. Devices within a cluster cooperatively handle shared processing steps, exchange intermediate results, and then independently complete the remaining tasks. This model is lauded for its adaptability, scalability, and efficient resource utilization.

Irrespective of the chosen architecture, the collaborative AIGC process commences with the training of high-quality AIGC models on robust computing platforms using extensive datasets. Post-training, these models are distributed to both edge servers and user devices to facilitate efficient task execution. Users initiate this process by submitting requests that describe their desired AIGC content. The system strategically analyzes these requests to optimize resource distribution and performance, employing a knowledge graph for semantic analysis to identify and group similar tasks. This facilitates the customization of shared processing steps for each group, thereby enhancing overall efficiency. The knowledge graph is dynamically updated to accommodate new tasks and adapt to user reclustering needs effectively.

For groups with analogous tasks, shared processing steps are executed on a central server using any relevant prompt from the group. Intermediate results are then forwarded to edge devices for subsequent processing. Ultimately, user devices receive these intermediate results and perform the final steps specific to their requests. This localized processing approach conserves energy, safeguards privacy, and empowers users to efficiently generate the AIGC content they desire. By distributing the processing load between a central server and user devices, this collaborative approach not only balances the workload but also minimizes latency, ensuring the generation of high-quality content.

\subsection{FL for Distributed Training of Large-scale AIGC Models}
The study in \cite{AIGC7} introduces a FL technique tailored for training large-scale AIGC models using massive datasets. This approach permits distributed clients to collaboratively train the model while retaining all training data locally, thereby preserving privacy.

This study investigates FL as a method to train high-capacity AIGC models, with a specific focus on the stable diffusion model, renowned for its high-quality image generation capabilities. FL-based methods enable distributed clients to collaboratively train the model while maintaining the privacy of their data. However, applying FL to AIGC models presents several challenges. For instance, in conventional FL, where each client trains the entire model, the computational demand can be high for clients with limited resources. Furthermore, when clients have varying data sizes and computational capabilities, the overall training process, particularly the convergence rate, can be adversely affected.

Parallel FL provides a degree of privacy protection but necessitates that all participants train the full model, which can be computationally intensive. On the other hand, the split learning approach within FL seeks to mitigate the computational load on resource-constrained clients by allowing them to take part in the training process. This method, however, requires a distinct separation of data roles among participants: one group holds all the training input samples, and another possesses only the corresponding labels. This requirement renders split learning inappropriate for certain AIGC models where local training at each client's site needs both the input data and its labels for effective backpropagation.

Given these limitations, the study proposes an optimized FL approach for fine-tuning the stable diffusion model, incorporating a D2D model-sharing technique to expedite the training process. This process unfolds in several steps. Specifically, in in initialization step, the parameter server selects a client and transmits the current model weights. In local training, the chosen client updates the model using its data and then exits the active pool. During round completion, if all clients have participated, the round ends; otherwise, the active client selects another from the remaining pool to continue training. In training completion, after a predetermined number of rounds, the base model and the refined weights merge to form a personalized model capable of generating customized content efficiently.
This federated approach leverages network infrastructure to enable seamless communication among clients, facilitating collaborative model training while maintaining data privacy. This distributed training relies on adequate network bandwidth to handle model weight exchanges and client coordination effectively.

\subsection{Securing SemCom in Virtual Transportation Networks With Blockchain}
Virtual transportation networks in the metaverse challenge virtual service providers (VSPs) with data collection, transmission efficiency, and security. A recent study \cite{AIGC9} proposes a framework integrating AIGC, blockchain technology, and SemCom to address these concerns. This framework facilitates seamless interactions between the physical and virtual worlds. However, a security vulnerability exists – attackers could manipulate semantic data, altering its meaning before it reaches the blockchain.

To address this, the framework incorporates zero-knowledge proofs (ZKPs). ZKPs allow secure processing of semantic data while guaranteeing the validity of any transformations, without revealing the actual data. Edge devices (like smartphones) capture real-world images and convert them into semantic data. VSPs interpret this data and use AIGC services to render corresponding visuals within the metaverse. To distinguish manipulated data, edge devices apply spatial transformations (e.g., blurring) to the authentic data. ZKPs then ensure the blurred data originates from legitimate transformations. Edge devices leverage a security parameter and a program to generate a shared secret string containing keys for both performing the computations (evaluation key) and verifying the results (verification key). The evaluation key allows edge devices to generate a zero-proof demonstrating the relationship between the original and transformed data, without revealing the data itself. The verification process is handled jointly by the blockchain, edge devices, and VSPs. All parties can query the verification results stored on the blockchain, fostering trust in a decentralized manner. The verification key, stored as a smart contract input, determines whether to accept or reject the proof based on the provided outputs. As shown in \cite{AIGC9}, this defense mechanism satisfies completeness, soundness, and zero-knowledge.

\section{Challenges and Future Studies}\label{Challeng}

As discussed in Sections from~\ref{AIGCOpt} to~\ref{xGSecurityOpt}, xG wireless communication networks stand to benefit significantly from the adoption of GAI frameworks. Specifically, based on Tables~\ref{tab_ISAC}, ~\ref{tab_SemCom}, ~\ref{tab_Security}, GAI models offer the potential to revolutionize mobile networks by fostering \emph{enhanced efficiency}, encompassing both energy efficiency and scalability.  They can also promote network \emph{sustainability} through improved resource management and reduced energy consumption. Additionally, GAI facilitates dynamic \emph{reconfigurability}, enabling the network to adapt to changing demands and optimize performance in real-time.
GAI holds promise for improving the \emph{reliability and accuracy} of wireless communication.  Through real-time prediction and control of network elements, GAI empowers proactive network management, leading to improved \emph{decision-making} capabilities.

However, as discussed in Section~\ref{Complexity}, GAI models may have highly demanding training requirements. Their increasing complexity necessitates high-performance computing platforms to address these needs. These platforms typically offer features like \emph{scalable hardware} with powerful GPUs or tensor processing units (TPUs), \emph{enhanced networking} for efficient data transfer, and specialized \emph{efficient memory management and software libraries} optimized for generative AI tasks. \emph{Cloud-based infrastructure} options further enhance the development and training experience. 

While high-performance platforms are essential for training powerful GAI models, ensuring their efficient deployment within resource-constrained wireless networks requires further optimization. This optimization goes beyond just the hardware. Researchers are actively exploring techniques that address the trade-off between model accuracy and computational efficiency. This includes \emph{designing GAI model architectures} specifically the ones for wireless networking tasks by considering limitations on devices such as smartphones and BSs. Additionally, \emph{training on high-quality data} that reflects real-world network scenarios is crucial. Techniques like data augmentation can further enrich this data and potentially reduce training costs. Finally, advancements in \emph{quantization and pruning} aim to decrease the number of parameters or precision requirements in GAI models.  These optimization strategies bridge the gap between powerful GAI models and their practical application within next-generation wireless networks. The details are given below.

\subsection{Trade-off Between Model Accuracy and Computational Efficiency}
GAI paves the way for the development of intelligent and adaptable xG wireless communication. However, the trade-off between accuracy and computational efficiency remains a major challenge in applying GAI for network optimization. To address this challenge, several key aspects can be explored. The details are given in what follows.

\subsubsection{Designing GAI model architectures}
As discussed, different GAI models, including GANs, GDMs, and GFlowNets, are currently used to generate high-fidelity data for networking tasks and optimization. However, standard accuracy metrics alone may not be sufficient. To capture the efficiency aspects of the chosen models, alternative metrics that encompass both the quality of generated data (realism and usefulness) and the computational resources required for training and inference should be considered. Additionally, researchers are exploring the benefits of using pre-trained GAI models or transferring knowledge from larger, more accurate models to smaller, more efficient ones \cite{KD1, KD2}. 
Exploring \emph{collaborative learning} approaches is another ongoing research field. This involves utilizing multiple GAI models, each specializing in specific aspects of network data generation. By combining their strengths, these models could achieve high accuracy with improved overall efficiency.

\subsubsection{High-quality training data}
Current research for gathering high-quality data to train the GAI model for optimizing the xG wireless networks focuses on network traffic data, leveraging network simulations, and potentially even utilizing synthetic data generation techniques \cite{DataTrain1, DataTrain6}. Efficient and scalable methods for labeling network data, such as crowdsourcing or semi-supervised learning techniques, normalization, and outlier removal \cite{DataTrain4, DataTrain5, DataTrain6} are explored. These methods can accurately reflect the diverse scenarios and complexities of real-world networks.

While generic network data is valuable, ongoing research should place a greater emphasis on \emph{domain-specific data collection}. For example, data for congestion prediction might differ from data needed for resource allocation optimization. Therefore, the GAI model should learn the specific patterns relevant to the desired network behavior. Additionally, traditional data collection might be resource-intensive. Ongoing research explores techniques like \emph{active/adaptive learning} \cite{DataTrain7, DataTrain8}, where the GAI model itself identifies the most informative data points for further collection or annotation. Finally, network data often contains sensitive information. \emph{FL-based techniques} \cite{DataTrain2, DataTrain3} can be explored to leverage the benefits of larger, diverse datasets while maintaining data privacy.

\subsubsection{Quantization and pruning}
For efficient GAI models for networking, quantization and pruning techniques would play a crucial role. Various quantization methods to reduce the precision of weights and activations within GAI models involve techniques like post-training quantization and quantization-aware training. The former involves quantizing a pre-trained model to a lower bit-width representation (e.g., from 32-bit floats to 8-bit integers). While the latter trains the model from scratch with lower precision weights and activations. On the other hand, different pruning strategies remove redundant connections within the GAI model architecture. 
Accordingly, the focus of current research is the trade-off between the level of quantization/pruning \cite{Quant1, Quant2, Prun1, Prun3} and the accuracy of the generated network data.

It is noteworthy that  quantized models need to be compatible with the target hardware platforms used for network applications. Therefore, designing quantization techniques which are compatible with network processing units (NPUs) or field-programmable gate arrays (FPGAs)\footnote{FPGAs are semiconductor devices that offer both flexibility and speed. They are built from tiny building blocks called logic blocks, which can be programmed to do different jobs. Unlike regular processors that follow fixed instructions, FPGAs can be rewired to connect these blocks in different ways. This lets data and control signals move freely around the chip, making FPGAs useful for tackling complex tasks that can change over time.} enhances the efficiency. Additionally, ongoing research can focus on jointly optimize both the GAI model architecture and the quantization/pruning techniques. This could involve co-designing the model with specific considerations for efficient quantization or incorporating pruning strategies within the training process.  

\subsection{High-performance Platform for GAI}
GAI models are a subclass of DL, and rely heavily on the capabilities of the underlying DL platform. In other words, optimizing the DL platform as the base of GAI, is critical to unlocking GAI's full potential and achieving desired results. The details are discussed below.

\subsubsection{Scalable hardware}
Training DNNs used in GAI models requires specialized hardware (HW) accelerators to efficiently handle the heavy computational demands of training algorithms. These accelerators include GPU-based, TPUs, FPGA-based, and application-specific integrated circuit (ASIC)-based designs\footnote{ASICs are built for a targeted task, letting them to achieve peak performance, lower power consumption, and smaller size compared to their flexible counterparts. This specialization makes ASICs ideal for high-volume production in devices like smartphones and AI accelerators.}.

For choosing a GPU, theoretical performance metrics such as peak performance and memory bandwidth are a starting point, but do not capture the whole picture. To assess real-world performance for a specific task, reference benchmarks like MLPerf \cite{AccDL1} can be used. Different vendors like AMD, Intel, and NVIDIA propose different architectures. For instance, NVIDIA DGX systems specifically designed for DL workloads using GPUs, combine high-performance CPUs with multiple GPUs interconnected through NVIDIA's high-speed NVLink technology \cite{nvidia_dgx1, nvidia_dgx2}. Additionally, TPUs have become prominent due to the prevalence of matrix multiplication operations in DL, especially for large CNNs. Google pioneered TPUs and keeps improving their performance \cite{TPU1,TPU2, TPU3}.

FPGAs are also popular for hardware acceleration owing to their ability to be reconfigured. For instance, Microsoft's Project Brainwave adopts FPGAs to exploit their reconfigurability for handling various deep learning models \cite{FPGA5}.
FPGA-based accelerators often leverage a heterogeneous architecture to meet the diverse computational needs of DL workloads. The key components include a \emph{general-purpose processor}, \emph{dedicated computational modules}, and a \emph{custom memory architecture}. The general-purpose processor handles software tasks like model loading and pre-processing \cite{FPGA1, FPGA2}. Meanwhile, dedicated computational modules are tailored for specific DL operations, such as convolutions, de-convolutions, and pooling. These modules exploit the inherent parallelism of FPGAs to achieve high efficiency \cite{FPGA3, FPGA4}. Additionally, data movement between the processing units and external memory is optimized by the custom memory architecture, minimizing latency and maximizing throughput. Deeply pipelined multi-FPGA designs are explored to handle large models, with ongoing research on optimization \cite{FPGA6}.

For DL inference using ASIC-based accelerators, accelerating the computations within the data path is crucial. Key techniques to achieve this include: reduced-precision convolutions \cite{ASIC1} to improve computational efficiency, approximate multipliers \cite{ASIC2} that trade lower power consumption for a slight accuracy reduction, and bit-width reduction techniques \cite{ASIC3} that focus on reducing the number of bits used in multiplications. Additionally, neural processing units (NPUs) can be incorporated into ASIC-based accelerators to further improve performance and energy efficiency \cite{ASIC4}.  Single-chip NPUs offer advantages over CPUs and GPUs for DNN processing in both cloud and edge computing scenarios, due to their optimized data flow that minimizes memory access \cite{ASIC5}.

As discussed in \cite{Surnry_AccDL}, future paradigms in DL accelerator design envision the development of accelerators for sparse matrices \cite{AccDL2}, 3D-stacked processing-in-memory \cite{AccDL3}, in-memory computing \cite{AccDL4}, neuromorphic accelerators \cite{AccDL5}, and multi-chip modules \cite{AccDL6}. These DL accelerators will be employed across various computing systems deployed in xG wireless networks, ranging from ultra-low-power and resource-constrained devices at the edge to servers and data centers.

\subsubsection{Enhanced networking}
Studies in Section~\ref{NetForGAI} highlight how networking strategies can enhance GAI models by enabling secure and traceable data flow and communication \cite{AIGC4, AIGC9}, decentralized network management (e.g., blockchain \cite{AIGC4}), optimized resource allocation and reduced latency for improved user experience (e.g., device collaboration \cite{AIGC11}), scalability, privacy preservation, and improved efficiency for resource-constrained devices (e.g., federated learning \cite{AIGC7}).

Ongoing research and development in network technologies for high-performance GAI platforms needs to address several challenges: scalability and performance, security and privacy, resource constraints and heterogeneity, and integration and interoperability.

\textbf{Scalability and performance:} High-performance GAI platforms demand efficient communication with minimal latency. Techniques like blockchain in \cite{AIGC4}  and secure data transmission (e.g., ZKPs in \cite{AIGC9}) offer benefits but can introduce overhead. Optimizations or alternative approaches are needed to balance these trade-offs, while network infrastructure must scale to manage communication between a potentially large number of distributed processing units.

\textbf{Security and privacy:} High-performance GAI platforms require robust security to mitigate network vulnerabilities that could compromise training data or manipulate models. Additionally, they should ensure both security and efficiency. In other words, these measures need to be balanced to avoid sacrificing performance or introducing excessive latency. 

\textbf{Resource constraints and heterogeneity:} GAI platforms encompass devices with varying capabilities. Network protocols and resource allocation strategies need to adapt to this heterogeneity. Lightweight communication protocols and techniques are crucial for enabling the efficient participation of resource-constrained devices while optimizing the utilization of all available resources.

\textbf{Integration and interoperability:} For broader adoption, GAI platforms need standardized network protocols and APIs to simplify integration between diverse hardware and software components. This interoperability must seamlessly coexist with existing network infrastructure, even as GAI platforms introduce new functionalities and security measures.

\subsubsection{Memory management and software libraries}
Existing libraries like TensorFlow and PyTorch provide building blocks for GAI development, e.g., \cite{tenGAN, pyGAN}. However, research can focus on specialized libraries optimized for specific GAI tasks (e.g., image generation, natural language processing) or hardware platforms. For memory management, training methods such as automatic mixed-precision (AMP) training and gradient accumulation (GA) are employed. The frameworks offering built-in AMP functionalities automatically analyze the model and identify parts where lower precision calculations can be used without significantly impacting accuracy \cite{CompMem1, CompMem2}. While GA accumulates the gradients calculated for each mini-batch over several iterations before performing a single update. This technique effectively simulates training with larger batch size, even if GPU memory constraints limit data processing in smaller chunks \cite{CompMem3, CompMem4}. Further research directions can be given as follows.

\textbf{Memory management for emerging architectures:}  Neuromorphic computing, inspired by the human brain's structure and function, utilizes artificial neurons and synapses for parallel processing with potentially lower power consumption. To effectively leverage these emerging architectures in GAI platforms, memory management becomes crucial. This area necessitates exploring techniques like sparse data structures (e.g., compressed sparse row format) and model compression (e.g., pruning, quantization) to reduce memory footprint. Additionally, in-memory computing techniques hold promise for further efficiency gains.

\textbf{Heterogeneous memory management:} In heterogeneous computing environments often used in GAI platforms, it is crucial to develop strategies for efficient memory management. This may involve optimizing data movement between CPUs, GPUs, TPUs, and potentially other hardware components.  

\textbf{Domain-specific software libraries:} To improve memory access patterns and reduce computational overhead, specialized software libraries should be developed. These libraries should be optimized for specific GAI tasks or application domains.

\textbf{Auto-tuning and optimization:} This method can include exploring  frameworks that can automatically select or configure memory management strategies and software libraries based on the specific GAI model, hardware platform, and user requirements.

Addressing the aforementioned research directions can create more efficient and scalable memory management solutions and software libraries for high-performance GAI platforms. It is noteworthy that standardization of these techniques would facilitate easier integration of different tools and frameworks within a GAI platform, enhancing platform flexibility and developer productivity.

\subsubsection{Cloud-based infrastructure}
Cloud computing has emerged as a game-changer for training and deploying GAI models \cite{AWSGAI1}. Cloud-based infrastructure offers several benefits for GAI models, including scalability, cost-effectiveness, collaboration, and flexibility. Specifically, these platforms provide on-demand access to vast computing resources (CPUs, GPUs, TPUs) that can be \emph{scaled} up or down based on training needs. Users only \emph{pay} for the resources they utilize. By providing a shared workspace for accessing training data, models, and code, cloud-based infrastructure facilitates \emph{collaboration} among researchers and developers, streamlining the development process and promoting faster innovation. Finally, in terms of \emph{flexibility}, these platforms offer a variety of pre-configured DL environments and tools, allowing users to choose the best fit for their specific needs and reducing the time and effort required to set up training infrastructure.

Some of the leading cloud platforms for GAI include Amazon SageMaker \cite{AWSGAI2}, Google Vertex AI \cite{GoogleGAI}, and Microsoft Azure Cognitive Services \cite{MSGAI}. These platforms offer a comprehensive suite of tools and resources specifically designed for training and deploying both DL and generative models. They seamlessly integrate with TensorFlow and other popular DL frameworks, streamlining the development process. Notably, while these platforms provide cloud resources for training custom generative models, they also offer pre-trained generative models as APIs for tasks like text generation, image manipulation, and speech synthesis \cite{MSGAI}.

The landscape of cloud-based platforms for GAI is constantly evolving. Cloud computing plays a crucial  role in GAI research areas such as resource optimization, security and privacy, FL for GAI, and automating model training pipelines. The details are given below.

\textbf{Resource optimization:} Optimizing resource utilization within the cloud environment are continuously explored. This includes developing algorithms for efficient task scheduling and data transfer, minimizing idle time and energy consumption.

\textbf{Security and privacy:} As discussed in Section~\ref{NetForGAI}, ongoing research in cloud-based GAI prioritizes the development of robust security and privacy-preserving techniques. This is crucial as concerns regarding the potential misuse of data generated by powerful GAI models continue to grow. These techniques will be essential for ensuring the safe and ethical training and deployment of generative models in cloud environments.

\textbf{FL for GAI}: FL techniques are particularly well-suited for training generative models on distributed datasets residing on different devices or cloud instances without directly sharing the data itself. This is crucial for scenarios where data privacy is paramount, such as in healthcare or finance. Ongoing research actively investigates how to adapt and improve FL-based approaches for training generative models efficiently across geographically distributed cloud environments.

\textbf{Automating model training pipelines:} In ML, a model training pipeline refers to a series of interconnected steps that automate the process of building, training, and deploying an ML model. Automating model training pipelines is an active research area in GAI, particularly focusing on tasks like hyperparameter tuning, model selection, and resource allocation. This automation makes the process more efficient and accessible to a broader range of users.

\section{Case Study: Diffusion-based CA, Load Balancing, and Backhauling in NTNs}\label{NTN}
In the preceding sections, we have discussed the powerful capabilities of GAI models in addressing the complexities of xG wireless networks. Motivated by the need for innovative solutions to optimize resource allocation and enhance network performance, this case study focuses on the practical application of diffusion-based GAI models for resource allocation in non-terrestrial networks (NTNs). Traditional methods such as optimization theory or game theory, while effective in certain scenarios, face significant limitations in NTNs due to the highly dynamic and complex nature of LEO satellite constellations. These methods often struggle with scalability, adaptability, and real-time decision-making in environments with constantly changing network conditions and heterogeneous resources.

Additionally, traditional ML-based methods such as RL techniques, though useful, can be limited in their ability to explore the action space thoroughly. GAI enhances RL by encouraging agents to explore more diverse actions and potential solutions through the generation of novel scenarios and actions. By employing GDMs and multi-agent DRL, we illustrate how GAI can overcome these challenges, optimizing spectrum utilization, improving data rates, and enhancing overall network efficiency in NTNs. This case study exemplifies the transformative potential of GAI, showcasing its ability to drive significant advancements in the performance and reliability of xG wireless networks through enhanced exploration and resource optimization.

\subsection{Background}
Future communication systems need extensive connectivity, reliable coverage, and strong backhaul links. xG wireless systems, including 6G networks, are looking to NTNs to improve global communication infrastructure. LEO satellite constellations like Starlink and Lightspeed aim to provide faster data and lower latency, but managing limited spectrum resources and avoiding interference among these satellites is a major challenge. CA is proposed as a solution to improve data rates, coverage, and resource utilization in LEO-based NTNs. CA combines multiple frequency bands to improve user experience and spectrum efficiency. In terrestrial networks, CA enhances throughput for UEs by allowing them to access extra bands. This case study proposes a resource management framework for LEO satellite (LEOS) networks that leverages CA and load balancing across these bands for improved network performance and spectrum utilization.

CA in NTNs presents major challenges for resource management. These challenges stem from the dynamic and unpredictable nature of the network, the diverse range of devices served, and the varying QoS requirements of different applications. While DRL has been explored for CA in dynamic 5G environments \cite{CA1}, its dependence on exploration-exploitation trade-offs can lead to suboptimal policies. To overcome this limitation and achieve better sub-optimal decisions regarding CC activation/deactivation for LEOS in these dynamic settings, we propose a hybrid approach. This approach combines a GDM, specifically the DDPM, with a multi-agent DRL model. DDPM offers a probabilistic framework that explicitly handles uncertainty, aiding in effective decision-making and exploration \cite{AIGC-Survey1,DDPM}. 

\subsection{Network Model and Problem Formulation}
\begin{figure}[t!]
  	\centering
  	\includegraphics[width=\columnwidth]{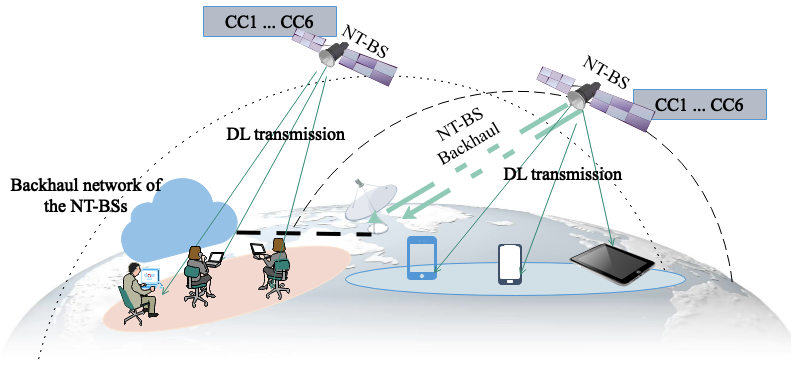}
  	\caption{System model of a LEOS-based non-terrestrial network with carrier aggregation technology.}
  	\label{NetModel}
  \end{figure}
Figure~\ref{NetModel} depicts an NTN consisting of LEOS  and a set of User Equipments (UEs), distributed across the Earth's surface. These LEOS connect to a ground station via a backhaul link, which acts as the gateway to the terrestrial internet infrastructure. The LEOS operate on pre-defined circular orbits, moving horizontally around the Earth. Additionally, they are equipped with CA technology and have on-board processing capabilities and decision-making algorithms that autonomously activate or deactivate CCs.  It is important to note that the utilization of CA in NTNs is contingent on the constellation design, allowing for the flexible use of CCs across different frequency bands such as L-band, S-band, and Ka-band. Our analysis considers a set of non-overlapping and non-orthogonal CCs, where each CC consists of multiple resource blocks (RBs). Each LEOS can activate multiple CCs to serve the UEs within its coverage area. LEOS's set of CCs includes both the primary CC (PCC) and secondary CCs (SCCs). The PCC serves as the main and always-active CC for each LEOS. In this case study, we employ a Round Robin algorithm to assign the PCC to a LEOS. SCCs, acting as auxiliary CCs, can be activated or deactivated for a LEOS to improve its achievable data rate.
The backhauling network utilizes the Ka-band, which is further divided into multiple subchannels.

The transmission model considers factors like active CCs for LEOS and path gain between LEOS and UEs. It defines a load factor to represent the portion of resources a LEOS dedicates to its covered UEs on a specific CC. This load factor is influenced by the amount of UE demand served on that CC. While backhaul model focuses on how data from LEOS is transmitted to the gateway on Earth via SCs in the Ka-band. The model considers factors like path gain between LEOS and the gateway, achievable rates, and backhaul capacity limitations. It defines a backhaul constraint that ensures the amount of data a LEOS transmits to users does not exceed its backhaul capacity.

In our main optimization problem for NTN resource management, we aim to find the optimal way to activate CCs, distribute traffic among them, and allocate backhaul resources for each LEOS. This problem balances two goals: maximizing overall network throughput (data transfer rate) and minimizing the total load on the LEOS across all CCs. Subject to the constraints of UE demand satisfaction and backhaul limitations, the problem can be stated as follows: 
\begin{subequations}\label{opt_Load_CA}
\begin{align}
\label{opt_Load_CA:a}
& \max
&& \text{Total achievable rate for the LEOs}
\\
\label{opt_Load_CA:aa}
& \min
&& \text{Total load on the LEOS across all CCs} \\
\label{opt_Load_CA:b}
& \mathrm{s.t.}
&& \text{Demand requirement for the UEs} \\
\label{opt_Load_CA:c}
& 
&& \text{Backhaul limitations} \\
& \mathrm{var.}
&& \text{Load factor over each CC}\\ 
& 
&& \text{SCs and CCs indicators}.
\end{align}
\end{subequations}
Due to the complexity for \eqref{opt_Load_CA}, the solution approach will involve a combination of DDPM and a multi-agent RL technique. The details are given in the next subsection.

\subsection{The Proposed Method}
To address the complexity of the problem formulated earlier, we  break it down into two smaller, interdependent problems. The first problem focuses on activating/deactivating CCs and assigning subchannels (SCs) for backhaul. The objective of this subproblem is to maximize the total achievable rate for LEOS \eqref{opt_Load_CA:a} while satisfying the backhaul constraints \eqref{opt_Load_CA:c}. The second problem deals with load balancing, optimizing how much each CC is utilized by the LEOS to serve its users. The objective of this problem is to minimize the total load on the LEOs across the CCs \eqref{opt_Load_CA:aa} while meeting the demand requirements of the UEs \eqref{opt_Load_CA:b}.

\begin{figure*}[t!] 
    \centering
    \includegraphics[width= 400pt]{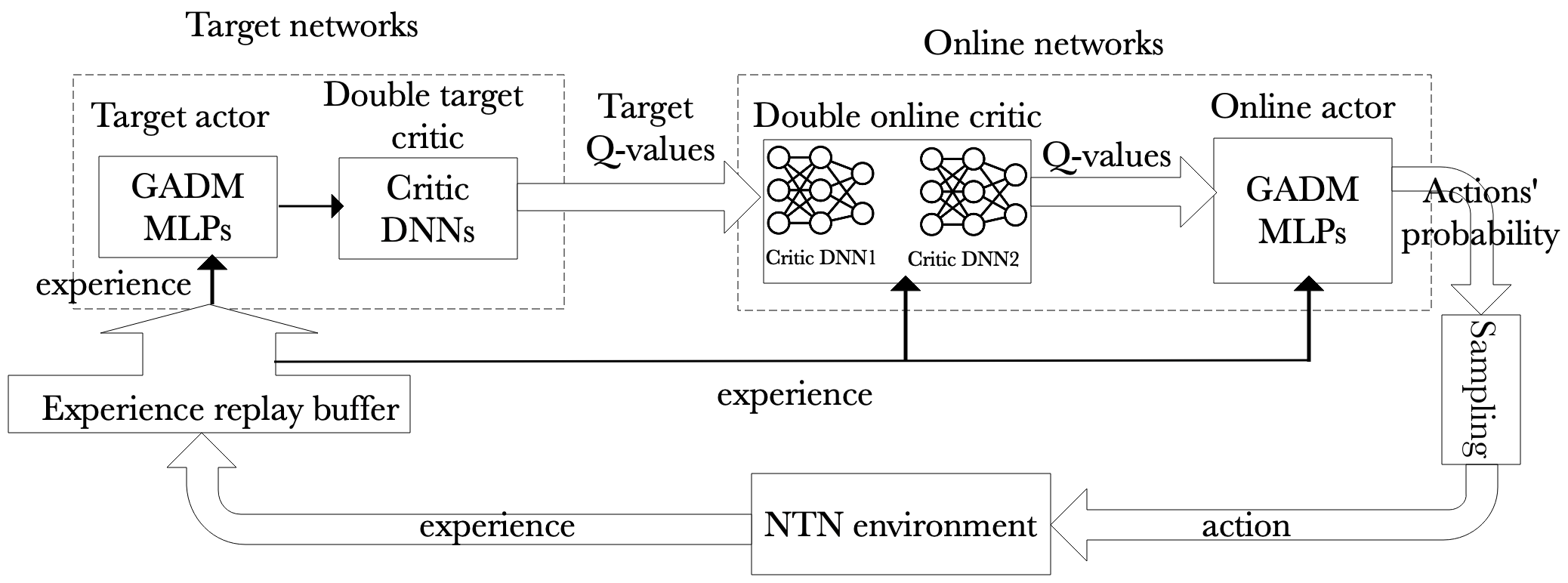} 
    \caption{Architecture and trajectory of the proposed diffusion-based method for CA and SC allocation.}
    \label{DA2CAB}
\end{figure*}

We model the joint CA and backhauling sub-problem as a multi-agent RL system. In this system, each LEO satellite acts as an agent, aiming to maximize its long-term reward, which is defined in terms of network-wide throughput and energy efficiency. The state space is defined by the backhaul constraint \eqref{opt_Load_CA:c}, and the action space is determined by the options to activate/deactivate a CC and assign the SCs. The high dimensionality of the state and action spaces for each agent hinders effective exploration using traditional RL methods. To overcome this limitation, we leverage the \textbf{GADM} algorithm (detailed in Algorithm~\ref{alg_GADM}) to guide the policy for selecting actions, such as activating/deactivating CCs and assigning SCs to the LEOs. {As illustrated in Fig.~\ref{DA2CAB}}, within our developed actor-critic architecture for solving the RL system, the actor network utilizes GADM to choose actions, while the critic networks evaluate their effectiveness. Both networks are trained iteratively, collecting experiences and updating their parameters to improve the policy over time. This approach enables effective exploration in the high-dimensional space and facilitates learning optimal resource allocation strategies for the LEO network.

The second sub-problem, load balancing, is formally expressed as a convex optimization problem with a unique optimal solution. It can be demonstrated that the optimality for this problem is achieved when constraint \eqref{opt_Load_CA:b} holds with equality. Using this information, we derive an iterative update function to determine the optimal load factors across the CCs. This function considers user demands served by each LEO satellite and their signal-to-interference-plus-noise ratio (SINR) to calculate an updated load factor for each LEO on each CC.

In the developed intelligent joint CA, load balancing and backhauling (IJCALB) algorithm, the previously described intelligent and iterative schemes operate together. Specifically, the IJCALB algorithm operates over multiple episodes, where each episode represents a cycle of learning and adjustment. The core of the algorithm involves two DNNs, an actor and a critic, for each LEOS. These networks interact to learn the optimal policy for allocating resources, i.e., CCs and SCs. During each episode, taking the current network state into account, a sub-optimal policy for a specific LEOS regarding CA and SC assignment is obtained. For each episode, once the LEOS has a proposed allocation policy, it calculates the load factor across its activated CCs through the distributed and iterative load balancing scheme. As the algorithm progresses through multiple episodes, the actor-critic networks in each LEOS learn and adjust. This process leads them towards an optimal policy for resource allocation – the best possible strategy for activating/deactivating CCs and assigning SCs. Correspondingly, the fixed point reached by the load updating also converges to the optimal load factor for that LEOS. The IJCALB algorithm allows each LEOS to learn and adapt its resource allocation strategy through a multi-episode training process, ultimately achieving an optimal configuration for the entire network.

\subsection{Numerical Results}
This section evaluates the performance of the proposed diffusion-based \textbf{IJCALB} algorithm against three RL-based algorithms for joint CA, load balancing, and backhauling  LEO-based networks. These algorithms are, UCB-based JCALB (UJCALB), DDQN-based JCALB (DJCALB), and diffusion UCB-based JCALB (DUJCALB). In \textbf{UJCALB} algorithm, upper confidence bound (UCB) strategy is used to handle the discrete nature of the CA and SC allocation problem. UCB encourages exploration of less-tested options while balancing estimated rewards. \textbf{DJCALB} leverages the double depp Q-Network (DDQN) algorithm in multi-agent RL for handling CA and backhauling. DDQN is a well-established technique for solving multi-agent RL problems. \textbf{DUJCALB} combines the UCB strategy with the GADM algorithm to provide additional guidance for the UCB policy, potentially leading to improved performance.

\begin{table}[t!]
\centering
\caption{LEOS-based network parameters and training hyperparameters}
    \label{param}
\resizebox{\columnwidth}{!}{%
\begin{tabular}{|c|c||c|c|}
\hline
\textbf{Network parameters}      & \textbf{Value} & \textbf{Hyperparameters}                      & \textbf{Value} \\ \hline \hline
DL transmit power for LEOS        & $10$ Watt        & Diffusion step (T)                            & $5$              \\ \hline
Carrier setting &
  \begin{tabular}[c]{@{}c@{}}CC1-CC5 are in K-band $400$ MHz \\ BW and frequency of $26$ GHz,\end{tabular} &
  \begin{tabular}[c]{@{}c@{}}Learning rates for actor\\ and critic networks\end{tabular} &
  \begin{tabular}[c]{@{}c@{}}$10^{-4}$,\\ $10^{-3}$\end{tabular} \\ \hline
Orbit radius for LEOS             & $8$ km           & entropy regularization ($\zeta$) & $0.05$           \\ \hline
UE's max. demand & $1.5$ MHz        & Discount factor ($\lambda$)      & $0.95$           \\ \hline
Number of SCs                    & $6$              & Batch size and replay buffer size             & $64,~10^5$       \\ \hline
 velocity  of LEOS                 & $2230$ m/sec     & $T_{\mathrm{episod}}$, $T_{\mathrm{step}}$                             & $200,~200$       \\ \hline
Total number of UEs              & $400$            & Weight of soft update                         & $0.005$          \\ \hline
\end{tabular}%
}
\end{table}
To evaluate the performance of our proposed diffusion-based \textbf{IJCALB} algorithm, we consider a $60~\mathrm{km} \times 60~\mathrm{km}$ square area. We assume that $400$ UEs are uniformly scattered in the area, and the covering satellites have a circular orbit with an $8~\mathrm{km}$ radius and randomly selected origin. Each LEOS covers a set of UEs, and not all UEs are necessarily covered by the LEOS'. Each of the LEOS is moving at a velocity of $2230$ m/sec. For CA, we assume that the CCs are in the Ka-band. Additionally, the channel between the satellites and the gateway is divided into some SCs.  The network parameters and the hyperparameters employed by the diffusion-based GADM algorithm are given in \textbf{Table~\ref{param}}. 

\begin{figure*}[t!]
    \centering
    \begin{subfigure}[b]{0.37\textwidth} 
        \centering
        \includegraphics[width=\textwidth]{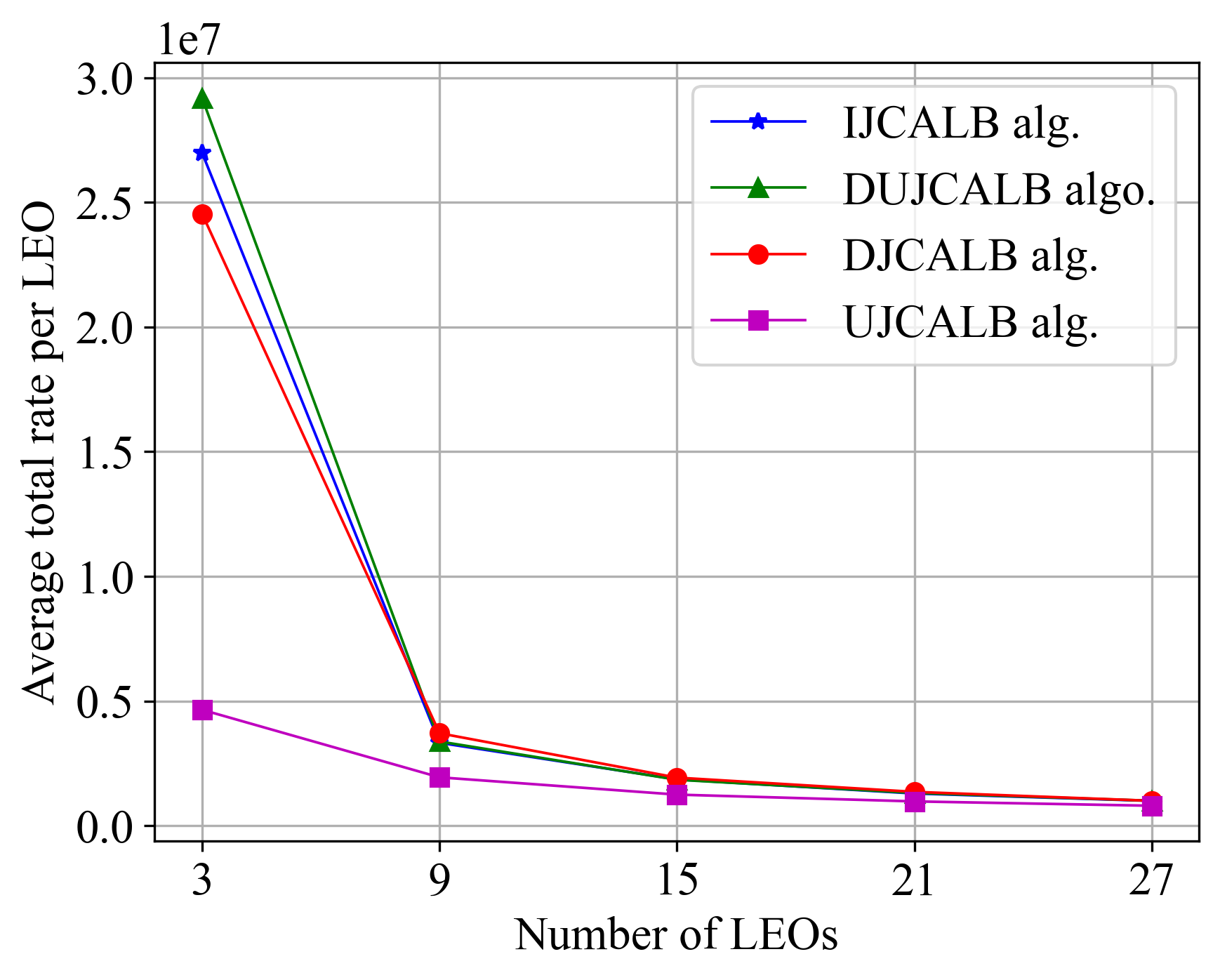}
        \caption{}
        \label{fig_AveRate}
    \end{subfigure}
    \hfill
    \begin{subfigure}[b]{0.37\textwidth} 
        \centering
        \includegraphics[width=\textwidth]{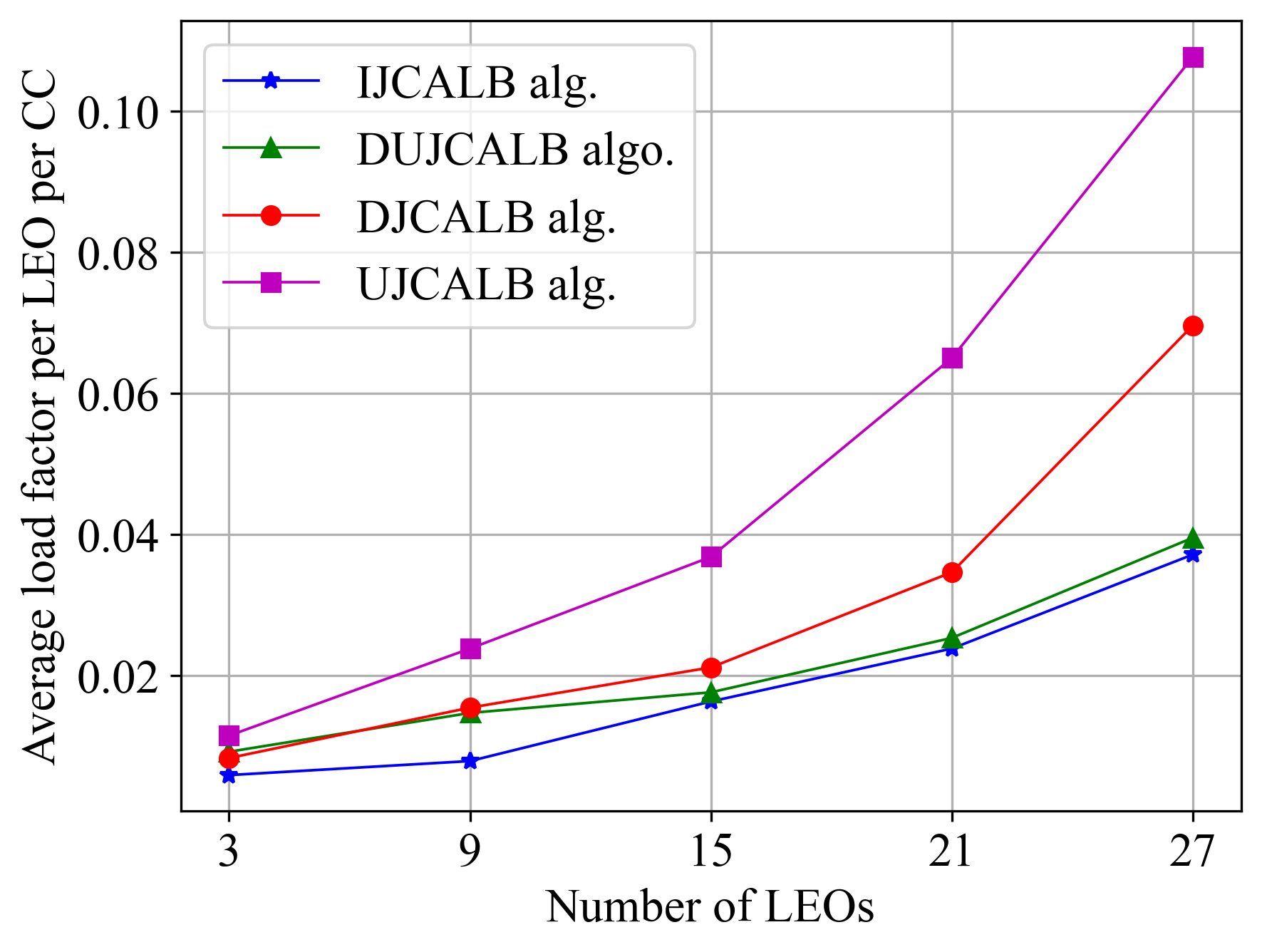}
        \caption{}
        \label{fig_AveLoad}
    \end{subfigure}
    \caption{Average achievable total rate per LEOS and average load factor for each LEOS per CC versus the number of LEOS.}
    \label{fig_Rate_Load}
\end{figure*}

\begin{figure*}[t!]
    \centering
    \begin{subfigure}[b]{0.37\textwidth}
        \centering
        \includegraphics[width=\textwidth]{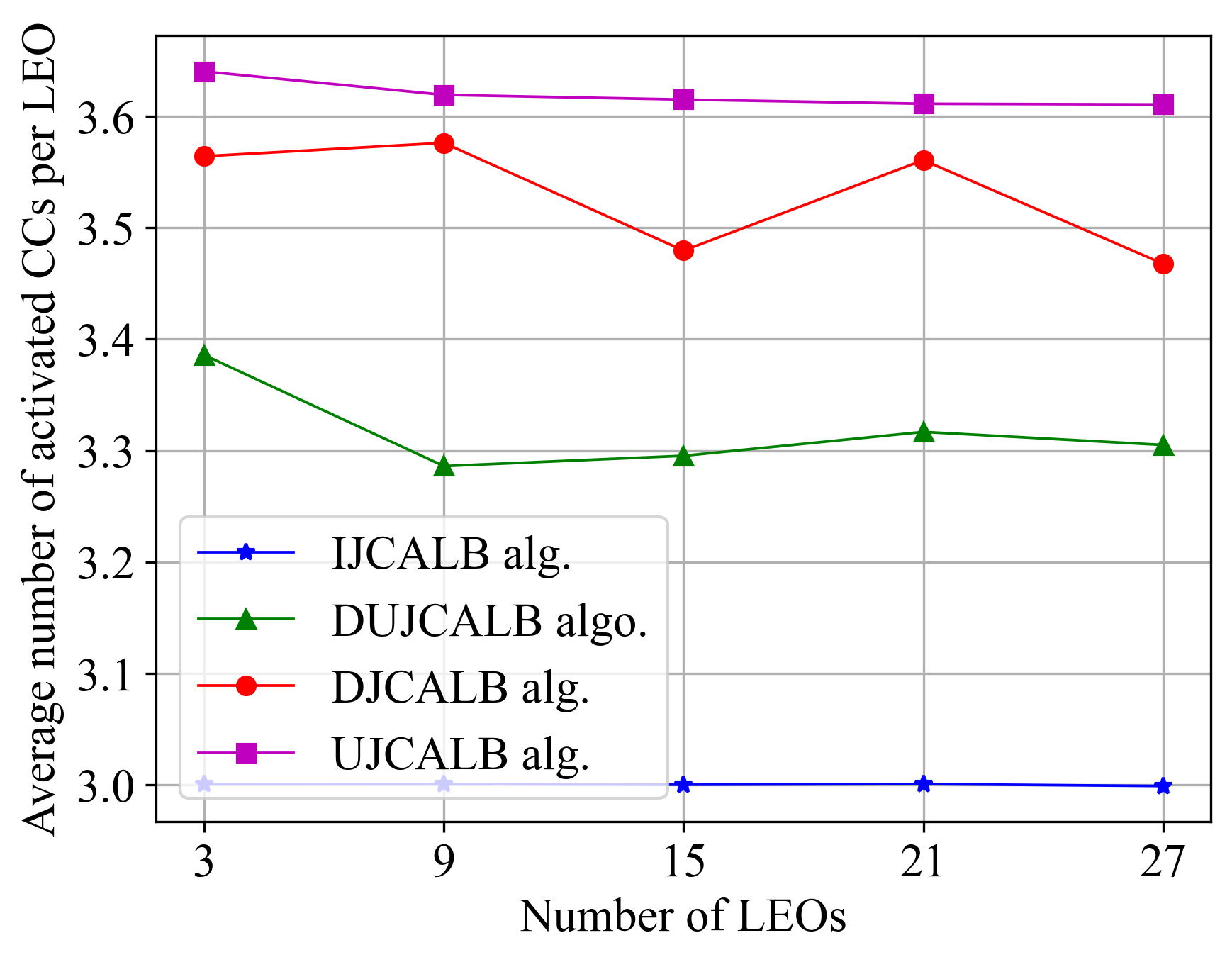}
        \caption{}
        \label{fig_AveCC}
    \end{subfigure}
    \hfill
    \begin{subfigure}[b]{0.37\textwidth}
        \centering
        \includegraphics[width=\textwidth]{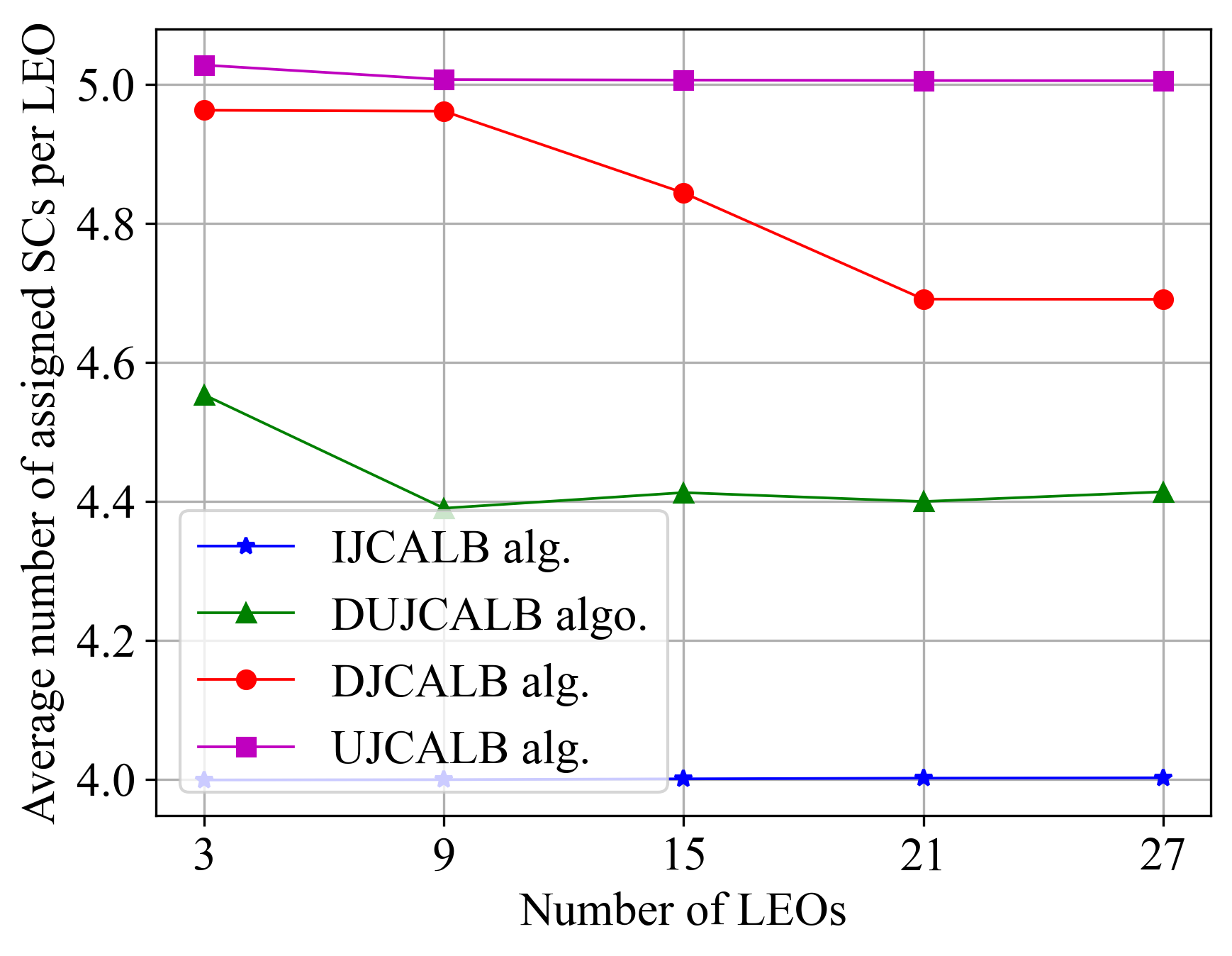}
        \caption{}
        \label{fig_AveSC}
    \end{subfigure}
    \caption{Average number of activated CC per LEOS and average number of assigned SCs per LEOS versus the number of LEOS.}
    \label{fig_CC_SC}
\end{figure*}
We evaluate the performance of the presented algorithms (IJCALB, DUJCALB, DJCALB, and UJCALB) across a varying number of LEOS in the network. The number of LEOS considered is $I \in \{3, 9, 15, 21, 27\}$. In \textbf{IJCALB} and \textbf{DUJCALB}, the policy for activating/deactivating the CCs and assigning the SCs to each LEOS is obtained through the diffusion-based GADM algorithm, while \textbf{DJCALB} employs DRL and traditional UCB is used by \textbf{UJCALB}. Figs.~\ref{fig_Rate_Load} and \ref{fig_CC_SC} illustrate the average total rate, average load factors over each activated CC, the average number of activated CCs, and the average number of assigned SCs per LEOS, respectively. 

{Fig.~\ref{fig_AveLoad}} illustrates that the load factor per activated CC per LEOS increases with the growing number of LEOS. A higher number of LEOS often indicates denser coverage areas, resulting in a greater number of served UEs. Consequently, there is an increased demand for spectrum resources from each CC, leading to a higher load factor for each LEOS. The increase in load factor per CC results in heightened interference between the UEs covered by LEOS' transmitting over the same CC. {Fig.~\ref{fig_AveRate}} illustrates that this interference can degrade the quality of received signals, thereby reducing the achievable data rate for the UEs and, consequently, for each LEOS. As shown in {Fig.~\ref{fig_AveLoad}}, the DDPM-based algorithms outperform the DDQN-based algorithm in terms of load factor. The DDPM-based policies employed in the \textbf{IJCALB} and \textbf{DUJCALB} algorithms enable them to estimate uncertainty. By incorporating this into the action selection process, these algorithms can make more informed decisions in dynamic environments. Thus, by employing the DDPM-based algorithms, CCs with better channel conditions would be activated for the LEOS', resulting in a lower load factor for the LEOS' over the activated CCs. In contrast, the DDQN-based \textbf{DJCALB} algorithm may struggle to effectively handle uncertainty. The inferior performance of the traditional UCB-based \textbf{UJCAB} algorithm can be attributed to its limitation in exploration, which would prevent it from discovering optimal policies for activating/deactivating the CCs and assigning SCs compared to the DDQN and the DDPM algorithms.

{Fig.~\ref{fig_CC_SC}} illustrates that a lower number of CCs and SCs are allocated to the LEOS' when using the two DDPM-based algorithms, i.e., \textbf{IJCALB} and \textbf{DUJCALB} algorithms. Based on the objectives of the subproblems, the DDPM-based algorithms aim to maximize the total rate for a LEOS'. Additionally, the state space is defined in terms of the achievable rate for the covered UEs, their QoS demand, and the dynamic backhaul constraint for the LEOS'. By leveraging the DDPM algorithm's ability to estimate uncertainty, the algorithm can select the CCs and SCs that are most likely to provide high-quality and reliable connections for the LEOS' and their covered UEs. As a result, a lower number of activated CCs and assigned SCs with higher achievable rates are obtained. It can also be inferred that the DDPM-based algorithms would be more energy-efficient due to the lower activation/deactivation of the CCs. It is worth mentioning that the \textbf{IJCALB} algorithm exhibits better performance compared to the \textbf{DUJCALB} algorithm due to its integration of DDPM into an actor and critic model.

\section{Conclusion}\label{Conc}
We have explored the potential of GAI models, such as GANs and GDMs, in addressing the critical challenges of optimizing xG wireless networks. By considering key technologies in 6G including  mobile AIGC, SemCom, ISAC, and security communication, we have reviewed how GAI's ability to learn complex network dynamics, generate diverse scenarios, and adapt to changing conditions can offer significant advantages over traditional optimization techniques. This survey has also explored how network infrastructure can be improved to utilize GAI-based models more effectively. The integration of GAI with existing AI-based network optimization models has been discussed in a case study, further expanding its capabilities. While challenges remain in terms of model complexity and training data requirements, ongoing research in distributed learning, edge computing, and on-device processing holds great promise for overcoming these issues.

\bibliographystyle{IEEEtran}
\bibliography{TutorialBib}

\end{document}